\documentclass{aa}
\usepackage{txfonts}
\usepackage{graphicx}
\usepackage{lscape}
\usepackage{natbib}
\bibpunct{(}{)}{;}{a}{}{,} 
\newcommand{\little}{\fontsize{8}{9}\selectfont} 
\begin{document}
\title{Molecules in bipolar proto-planetary nebulae
  \thanks{Figures~\ref{fig16594-1}-\ref{fig17150} are only available
  in electronic form via {\tt http://www.edpsciences.org}}
}
\author{P.~M.~Woods\inst{1,2}, L.-\AA.~Nyman\inst{2,3}, F.~L.~Sch\"
oier\inst{4}, A.~A.~Zijlstra\inst{1}, T.~J.~Millar\inst{1},
H.~Olofsson\inst{4}}
\titlerunning{Molecules in bipolar PPNe}
\authorrunning{Woods et al.}
\offprints{TJM,\\ ~\email{Tom.Millar@umist.ac.uk}}
\institute{Department of Physics, UMIST, P.O. Box 88, Manchester M60
1QD, UK.  
\and European Southern Observatory, Alonso de Cordova 3107,
Casilla 19001, Santiago 19, Chile.
\and
Onsala Space Observatory, SE-439 92 Onsala, Sweden.
\and
Stockholm Observatory, AlbaNova, SE-106 91 Stockholm, Sweden.
}
\date{Received <date> / Accepted <date>}
\abstract{Two bipolar proto-planetary nebulae, \object{IRAS16594-4656}
and \object{IRAS17150-3224}, have been detected in various molecular
lines, namely CO, $^{13}$CO, HCN and CN, and remain undetected in
several other species. CO($J$=2-1) and CO($J$=3-2) line profiles are
compared to new spectra of similar PPN candidates, previously
undetected in CO($J$=2-1): \object{CPD-53\degr5736},
\object{IRAS17106-3046}, \object{IRAS17245-3951} and
\object{IRAS17441-2411}. CO($J$=2-1) maps of \object{IRAS16594-4656}
and \object{IRAS17150-3224} show that both PPNe have little separation
between blue, centre and red-shifted emission, and also that the
CO($J$=2-1) emission is of a similar size to the telescope beam.
Fractional abundances of all detected molecules (except CO) are
calculated using the results of CO line modelling and a simple
photodissociation model. For those species not detected, upper limits
are derived. Comparisons between these fractional abundances and those
of other PPNe show that \object{IRAS16594-4656} and
\object{IRAS17150-3224} are quite under-abundant when compared to
molecule-rich sources like \object{CRL618}, \object{CRL2688} and
\object{OH231.8+4.2}. As a reason for this deficit, the difference in
circumstellar envelope/torus density between the molecule-rich sources
and the molecule-poor sources is proposed, and supported by a chemical
model which follows the transition of a circumstellar envelope through
the AGB phase and into the PPN phase of evolution. The model includes
the effects of UV radiation, cosmic rays and also X-rays. Finally, the
post-AGB ages of these two objects ($200-400$~yr) are estimated
using CN/HCN and HCN/CO ratios and both ages are found to be in
agreement with previous figures cited in the literature,
\object{IRAS17150-3224} being the younger of the two PPNe.
 }
\maketitle

\section{Introduction}

The proto-planetary stage of evolution is one of the shortest in a
star's lifetime and this implies a scarcity of observable
examples.  Following the method of \citet{Olivier_etal2001}, the
number surface density of proto-planetary nebulae can be estimated to
be (0.41$^{+1.91}_{-0.34}$)\,kpc$^{-2}$. In comparsion, the density of
AGB stars is 15\,kpc$^{-2}$\ and of Main Sequence stars is
$\sim$2$\times$10$^6$\,kpc$^{-2}$ \citep{Olivier_etal2001}. Hence
knowledge about this phase is limited. Most of the current
understanding of PPNe is derived from a handful of objects; mainly the
carbon-rich sources \object{CRL618} and \object{CRL2688} and the
oxygen-rich object \object{OH231.8+4.2}. All three show strong
molecular lines, have axisymmetric structures and molecular tori or
disks. However, several hundred PPNe candidates have been identified
\citep[see][ and references therein]{Kwok1993}, some 34 are reasonably
well identified as PPNe \citep{Bujarrabal_etal2001}, and yet only a
handful show such molecular \textquotedblleft
richness\textquotedblright. The reason for this difference is far from
clear.

Many optical and infrared HST images have helped in the study of PPNe
\citep[e.g., see][ for details on the two PPNe described
here]{Su_etal2001,Hrivnak_etal2001,Hrivnak_etal1999,Kwok_etal1998},
although the depth to which many of the objects discovered have been
studied is minimal. Both molecule-rich PPNe and molecule-poor PPNe
appear similar in images -- the three PPNe mentioned above and both
\object{IRAS16594-4656} and \object{IRAS17150-3224} have some degree
of bipolarity, and a narrow waist. The beginnings of this bipolarity
are found in the late AGB phase, where the first signs of asphericity
are seen \citep[e.g.,][]{KastnerWeintraub1994}. The shaping of the
nebula continues under the influence of a fast superwind, according to
the Generalized Interacting Stellar Wind (GISW) model
\citep{Balick1987}. Other effects then play a part in the developing
morphology \citep[see the review by][]{BalickFrank2002}, and one which
may be important is the inertial confinement of the outflowing wind by
a circumstellar torus or disk \citep{CalvetPeimbert1983}. The degree
of collimation produced by this disk would depend on the mass of the
disk, as well as the momentum involved in the high-speed outflow, and
other, geometrical effects. This change in morphology, from something
approximately spherical to something bipolar, or elliptical, occurs
very rapidly at the end of the AGB phase of evolution
\citep[e.g.,][]{Kwok_etal1996, Schmidt_etal2002}. The actual period of
transition is hard to quantify, and there are inherent difficulties in
estimates by dynamical means \citep{Zijlstra_etal2001}.

In this paper, two objects which have been imaged by the HST and show
strong CO emission are studied. Both objects, \object{IRAS16594-4656}
and \object{IRAS17150-3224} are only detected in a handful of species,
including HCN and CO. These are compared to other PPNe newly observed
in HCN and CO - \object{CPD-53\degr5736}, \object{IRAS17106-3046},
\object{IRAS17245-3951} and \object{IRAS17441-2411}
(Sect.~\ref{COobs}). Furthermore, the molecular properties of these
two PPNe are studied by calculating fractional abundances from
SEST data, and comparing with similar objects (Sects.~\ref{abunds} \&
\ref{disc}). The comparative under-abundance of molecules of
\object{IRAS16594-4656} and \object{17150-3224} is discussed in
Sect.~\ref{disc}, and arising hypotheses are confirmed by means of a
chemical model (Sect.~\ref{chemmodel}). Finally, individual
spectra are presented in Appendix~\ref{spectra}, which is only
available in electronic form.

\section{Observations}

\begin{table}
   \caption{Positions and LSR velocities for the PPNe under study.}
\label{stellardata}
   \begin{flushleft}
   \begin{tabular}{lcccc}
   \hline\hline
 IRAS No.    & 
 Other name           & 
   \multicolumn{2}{c}{J2000 Co-ords.}                & 
  {v$_{\rm LSR}$}  \\
             &                           & 
              \multicolumn{1}{c}{[h:m:s]} & 
               \multicolumn{1}{c}{[\degr:\arcmin:\arcsec]} & 
                \multicolumn{1}{c}{[km\,s$^{-1}$] }\\
    \hline
\object{14488-5405} & \object{CPD-53\degr5736} & 14:52:28.7 & $-$54:17:43 & $-10$\\
\object{16594-4656} & ---                       & 17:03:09.7 & $-$47:00:28 & $-25$\\
\object{17106-3046} & ---                   & 17:13:51.7 & $-$30:49:40 & $0$\\
\object{17150-3224} & \object{AFGL6815S}    & 17:18:20.0 & $-$32:27:20 & $+15$\\
\object{17245-3951} & \object{OH348.8-2.8}  & 17:28:04.8 & $-$39:53:44 & $0$\\
\object{17441-2411} & \object{AFGL5385}     & 17:47:10.3 & $-$24:12:54 & $+110$\\
  \hline
   \end{tabular}
\end{flushleft}
\end{table}

\begin{table}
\begin{flushleft}
\caption{SEST FWHM beam widths and efficiencies at selected
frequencies.}
\label{effic}
\begin{tabular}{c cc}
\hline\hline
Frequency & FWHM & $\eta_{\rm mb}$ \\
{\rm [GHz]} & [$\arcsec$] & \\
\hline
\phantom{0}86&	57	&	0.75    \\
100	&	51	&	0.73	\\
115	&	45	&	0.70	\\
147     &       34      &       0.66    \\		
230	&	23	&	0.50	\\				
265	&	21	&	0.42	\\
345     &       15      &       0.25    \\				
\hline
\end{tabular}
\end{flushleft}
\end{table}

The observations were carried out between 1998 and 2002 with the
Swedish-ESO Submillimetre Telescope (SEST), situated on La Silla,
Chile. SIS receivers were used at 0.8, 1.3, 2 and 3\,mm.  During this
period the SEST operated three acousto-optical spectrometers: one high
resolution spectrometer, with a bandwidth of 86\,MHz and a channel
separation of 43\,kHz, and two wideband spectrometers with bandwidths
of about 1\,GHz and a channel separation of 0.7\,MHz. Typical system
temperatures above the atmosphere ranged between 150 and 800\,K,
depending on frequency and elevation.

Most observations were carried out with the dual beam switching method
which gives very flat baselines by placing the source alternately in
two beams. Beam separation was 11\farcm5 in azimuth. CO maps of
\object{IRAS17150-3224} were taken using the position switching
method, to minimise interference from interstellar CO emission
lines. The OFF position was chosen to be (-10\arcmin, -11\arcmin) from
the ON position. Map spacing in all cases was 11\arcsec, which is
approximately half a beamwidth at 230\,GHz. Calibration was performed
with the standard chopper-wheel technique, and has an uncertainty of
approximately 20\%\ \citep[see ][ for details]{Schoeier01}. J2000
positions of both sources are given in Table~\ref{stellardata}.

Intensity scales of spectra in this paper are given in main-beam
brightness temperature, which is the corrected antenna temperature
divided by the main-beam efficiency ($\eta_{\rm mb}$). Beam
efficiencies and FWHM beam widths are summarised in Table~\ref{effic}.
Individual spectra are shown in Appendix~\ref{spectra}, and line
frequencies and integrated intensities are shown in
Table~\ref{intintens}.

Integrated intensities of lines which were not clearly detected are
calculated using the following expression,
\begin{eqnarray}
\label{uleqn}
I_v \leq 3\sigma \left(\sqrt{2}\
\sqrt{2\frac{v_{\mathrm{exp}}}{\Delta v_{\mathrm{res}}}}\right)\Delta v_{\mathrm{res}}
= 3\sigma (2n)^{1/2}\Delta v_{\mathrm{res}}.
\end{eqnarray}
where $\sigma$\ is the rms noise in the spectrum, $v_{\mathrm{exp}}$\
the expansion velocity, $\Delta v_{\mathrm{res}}$ the velocity
resolution of the spectrum and $n$ the number of channels covering the
line width. This gives a somewhat optimistic view of the integrated
intensity of the line, and hence any fractional abundance
calculation using this value will be strictly an upper limit. Values
of $I_v$\ calculated in this manner are given in Table~\ref{intintens}
and indicated by a \textquotedblleft less than\textquotedblright\
sign.

\begin{table*}
   \caption{Observed lines.}
\label{intintens}
   \begin{flushleft}
   \begin{tabular}{lll ccc ccc}
   \hline\hline
 & & &&\multicolumn{2}{c}{\object{IRAS16594-4656}}&&\multicolumn{2}{c}{\object{IRAS17150--3224}} \\
\cline{5-6}\cline{8-9}
Molecule &Transition&Frequency&& $T_{\mathrm{mb}}$ & $\int T_{\rm mb}{\rm dv}$ && $T_{\mathrm{mb}}$ & $\int T_{\rm mb}{\rm dv}$\\
 & & GHz && K & K\,km\,s$^{-1}$ && K & K\,km\,s$^{-1}$ \\
    \hline
C$_3$S & $J$=15-14 & 86.708 && $<$\,0.01 & $<$\,0.03 && --- & --- \\
SiO & $J$=2-1 & 86.847 && $<$\,0.01 & $<$\,0.03 && --- & --- \\
HN$^{13}$C & $J$=1-0 & 87.091 && $<$\,0.01 & $<$\,0.03 && --- & --- \\
C$_4$H & $^2\Pi_{3/2}$, 19/2-17/2 & 87.372 && $<$\,0.01 & $<$\,0.03 && --- & --- \\
C$_2$H & $N$=1-0 & 87.329 && $<$\,0.01 & $<$\,0.03 && --- & --- \\
HCN & $J$=1-0 & 88.632 && {\bf\phantom{$<$\,}0.02} & {\bf\phantom{$<$\,}0.47} && {\bf\phantom{$<$\,}0.02} & {\bf\phantom{$<$\,}0.54 }\\
C$_3$N & $N$=9-8, $J$=19/2-17/2 & 89.046 && $<$\,0.01 & $<$\,0.05 && --- & --- \\
C$_3$N & $N$=9-8, $J$=17/2-15/2 & 89.064 && $<$\,0.01 & $<$\,0.05 && --- & --- \\
HCO$^+$ & $J$=1-0 & 89.189 && $<$\,0.01 & $<$\,0.05 && --- & --- \\
HC$^{13}$CCN & $J$=10-9 & 90.593 && $<$\,0.01 & $<$\,0.04 && --- & --- \\
HCC$^{13}$CN & $J$=10-9 & 90.602 && $<$\,0.01 & $<$\,0.04 && --- & --- \\
HNC & $J$=1-0 & 90.664 && $<$\,0.01 & $<$\,0.04 && --- & --- \\
SiS & $J$=5-4 & 90.772 && $<$\,0.01 & $<$\,0.04 && --- & --- \\
HC$_3$N & $J$=10-9 & 90.979 && $<$\,0.01 & $<$\,0.04 && --- & --- \\
CH$_3$CN & 5(1)-4(1) & 91.985 && $<$\,0.01 & $<$\,0.05 && --- & --- \\
HC$^{13}$CCN & $J$=12-11 & 108.711 && $<$\,0.01 & $<$\,0.06 && --- & --- \\
HCC$^{13}$CN & $J$=12-11 & 108.721 && $<$\,0.01 & $<$\,0.06 && --- & --- \\
$^{13}$CN & $N$=1-0 & 108.780 && $<$\,0.01 & $<$\,0.06 && --- & --- \\
C$_3$N & $N$=11-10, $J$=23/2-21/2 & 108.834 && $<$\,0.01 & $<$\,0.06 && --- & --- \\
C$_3$N & $N$=11-10, $J$=21/2-19/2 & 108.853 && $<$\,0.01 & $<$\,0.06 && --- & --- \\
SiS & $J$=6-5 & 108.924 && $<$\,0.01 & $<$\,0.06 && --- & --- \\
HC$_3$N & $J$=12-11 & 109.174 && $<$\,0.01 & $<$\,0.06 && --- & --- \\
$^{13}$CO & $J$=1-0 & 110.201 && {\bf\phantom{$<$\,}0.04} & {\bf\phantom{$<$\,}0.06} && {\bf\phantom{$<$\,}0.04} & {\bf\phantom{$<$\,}0.75} \\
CH$_3$CN & 6(1)-5(1) & 110.381 && $<$\,0.01 & $<$\,0.11 && $<$\,0.01 & $<$\,0.09 \\
C$_4$H & $^2\Pi_{3/2}$ 23/2-21/2 & 113.266 && $<$\,0.01 & $<$\,0.09 && $<$\,0.01 & $<$\,0.06 \\
C$_2$S & 8(7)-9(8) & 113.410 && $<$\,0.01 & $<$\,0.09 && $<$\,0.01 & $<$\,0.06 \\
CN & $N$=1-0 & 113.491 && {\bf\phantom{$<$\,}0.05} & {\bf\phantom{$<$\,}3.63} && $<$\,0.01 & $<$\,0.06 \\
C$_4$H & $^2\Pi_{3/2}$ 21/2-19/2 & 115.217 && $<$\,0.07 & $<$\,0.76 && $<$\,0.04 & $<$\,0.45 \\
CO & $J$=1-0 & 115.271 && {\bf\phantom{$<$\,}0.61} & {\bf\phantom{$<$\,}1.78} && {\bf\phantom{$<$\,}0.25} & {\bf\phantom{$<$\,}5.05} \\
SiC$_2$ & $5_{0,5}$-$4_{0,4}$ & 115.382 && $<$\,0.07 & $<$\,0.76 && $<$\,0.04 & $<$\,0.45 \\
SiO & $J$=3-2 & 130.269 && $<$\,0.01 & $<$\,0.06 && --- & --- \\
H$_2$CO & $2_{1,2}$-$1_{1,1}$ & 140.840 && $<$\,0.01 & $<$\,0.07 && --- & --- \\
SiC$_2$ & $6_{2,5}$-$5_{2,4}$ & 140.920 && $<$\,0.01 & $<$\,0.07 && --- & --- \\
H$^{13}$CCCN & $J$=16-15 & 141.062 && $<$\,0.01 & $<$\,0.07 && --- & --- \\
CS & $J$=3-2 & 146.969 && $<$\,0.01 & $<$\,0.04 && --- & --- \\
CH$_3$CN & 8(0)-7(0) & 147.175 && $<$\,0.01 & $<$\,0.04 && --- & --- \\
$^{13}$CO & $J$=2-1 & 220.399 && {\bf\phantom{$<$\,}0.10} & {\bf\phantom{$<$\,}1.34} && {\bf\phantom{$<$\,}0.13} & {\bf\phantom{$<$\,}2.91} \\
CH$_3$CN & 12(0)-11(0) & 220.747 && $<$\,0.01 & $<$\,0.16 && $<$\,0.01 & $<$\,0.17 \\
CN & $N$=2-1 & 226.875 && {\bf\phantom{$<$\,}0.11} & {\bf\phantom{$<$\,}9.91} && $<$\,0.01 & $<$\,0.10 \\
CO & $J$=2-1 & 230.538 && {\bf\phantom{$<$\,}1.75} & {\bf\phantom{\~}32.60} && {\bf\phantom{$<$\,}0.65} & {\bf\phantom{\~}12.84} \\
CO$^+$ & $J$=2-1 & 235.790 && $<$\,0.01 & $<$\,0.12 && --- & --- \\
SiS & $J$=13-12 & 235.961 && $<$\,0.01 & $<$\,0.12 && --- & --- \\
CH$_3$CN & 13(0)-12(0) & 239.138 && $<$\,0.01 & $<$\,0.16 && --- & --- \\
CH$_3$C$_2$H & 14(0)-13(0) & 239.252 && $<$\,0.01 & $<$\,0.16 && --- & --- \\
C$_3$H$_2$ & $4_{4,1}$-$3_{3,0}$ & 265.759 && $<$\,0.02 & $<$\,0.36 && --- & --- \\
HCN & $J$=3-2 & 265.886 && {\bf\phantom{$<$\,}0.11} & {\bf\phantom{$<$\,}2.85} && --- & --- \\
H$^{13}$CN & $J$=4-3 & 345.340 && $<$\,0.04 & $<$\,0.80 && $<$\,0.06 & $<$\,1.28 \\
HC$_3$N & $J$=38-37 & 345.610 && $<$\,0.04 & $<$\,0.80 && $<$\,0.06 & $<$\,1.28 \\
CO & $J$=3-2 & 345.796 && {\bf\phantom{$<$\,}2.05} & {\bf\phantom{\~}41.20} && {\bf\phantom{$<$\,}0.78} & {\bf\phantom{\~}18.19} \\
  \hline
   \end{tabular}
\end{flushleft}
\end{table*}

\section{Sources}

The two PPNe presented in this paper are objects for which there are
currently few millimetre-wavelength spectra and little or no molecular
information. Both show bipolar morphology in HST images and also
bright CO emission, with interesting features. These two sources were
selected on the basis of previously published CO spectra
\citep{Loup_etal1990,Hu_etal1993}.

\subsection{\object{IRAS16594-4656}}

\object{IRAS16594-4656} is a proto-planetary nebula which lies almost
in the plane of the Galaxy \citep[$b=-3$\fdg3, ][]{VDSteeneVHoof2003}
and has a mixed chemistry. In its infrared spectrum there are strong
and rarely observed features at 12.6 and 13.4\,$\mu$m, thought to be
due to polycyclic aromatic hydrocarbons (PAHs) with a high degree of
hydrogenation, and further PAH features in the range 3-13\,$\mu$m
\citep{Garcia_etal1999}. An emission feature between 19 and
23\,$\mu$m, associated with C-rich PPNe by \citet{KwokHrivnak1989},
also suggests that this source is carbon rich. However, there are also
indications of crystalline silicates (pyroxenes), with weak features
around 34\,$\mu$m \citep{Garcia_etal1999}, and a 10\,$\mu$m silicate
feature \citep{Olnon_etal1986}. This suggests that until recently this
object was oxygen-rich, becoming carbon-rich shortly before the
transition to the post-AGB phase \citep{Garcia_etal1999}, which
occurred some 370 years ago \citep{VDV_etal1989}. This hypothesis was
strengthened by the lack of a detection in OH
\citep{Hekkert_etal1991,Silva_etal1993} and no SiO maser emission
\citep{Nyman_etal1998}. In evolved stars with a mixed chemistry
such as this, the silicates are found in a torus and the PAHs in the
polar flow \citep{Matsuura_etal2004}.

Several authors describe HST images of this PPN
\citep{Hrivnak_etal1999,Hrivnak_etal2000,Garcia_etal1999}, which show
the presence of a bright , B7-type \citep{VDSteene_etal2000} central
star, surrounded by a multiple--axis bipolar nebulosity with a complex
morphology. The nebula is seen at an intermediate inclination, rather
than edge-on. The optical size of the nebula is given as
5\arcsec$\times$11\arcsec \citep{Garcia_etal1999}.

The only CO spectrum of this source previously published is a CO
($J$=1-0) spectrum from \citet{Loup_etal1990}, who detected broad CO
emission corresponding to that from a circumstellar shell expanding
with a velocity of 16\,km\,s$^{-1}$.

\subsection{\object{IRAS17150-3224}}

\object{IRAS17150-3224} (\object{AFGL6815S}) is a young PPN which left
the AGB some 150 \citep{Hu_etal1993} to 210 years ago, and possibly
came from a high-mass progenitor \citep{Meixner_etal2002}. Note
however, that \citet{VDV_etal1989} give a dynamic timescale of 800\,yr
for this object. Structurally, it is a bipolar nebula, with a nearly
edge-on (82\degr) dusty torus \citep{Kwok_etal1996}, expanding at
11\,km\,s$^{-1}$ \citep{Weintraub_etal1998} around a G2-type central
star. The optical size of the nebula is approximately 12\arcsec\
$\times$\ 9\arcsec\
\citep{Hu_etal1993,Kwok_etal1996,Garcia-H_etal2002}. A
newly-discovered equatorial loop seen in near-infrared, but not
optical, images, is discussed by \citet{Su_etal2003}. A faint halo
(AGB mass-loss remnant) can be seen in images from
\citet{Kwok_etal1998}.

\object{IRAS17150-3224} shows molecular hydrogen emission, but not
hydrogen recombination, indicating that the degree of ionisation is
small \citep{Garcia-H_etal2002}. The emission is likely due to
shock-excitation, as is often seen in strongly bipolar nebulae. This
object is similar to \object{CRL2688} \citep{Sahai_etal1998} in that
it has H$_2$ emission, and a spectral type later than A
\citep[\object{CRL2688} has a spectral type of F2, ][]{CohenKuhi1977}.
\object{IRAS17150-3224} shows OH maser emission \citep{Hu_etal1994},
but not SiO maser emission \citep{Nyman_etal1998}. It also has been
detected in the 3.1\,$\mu$m water line \citep{VDV_etal1989}, but not
in H$_2$O maser emission \citep{ZuckLo1987}.

Previous CO observations are limited to the paper of
\citet{Hu_etal1993}, who present both CO ($J$=1-0) and ($J$=2-1)
spectra.

\section {CO observations}
\label {COobs}

   \begin{figure*}
   \sidecaption
   \includegraphics[width=12cm]{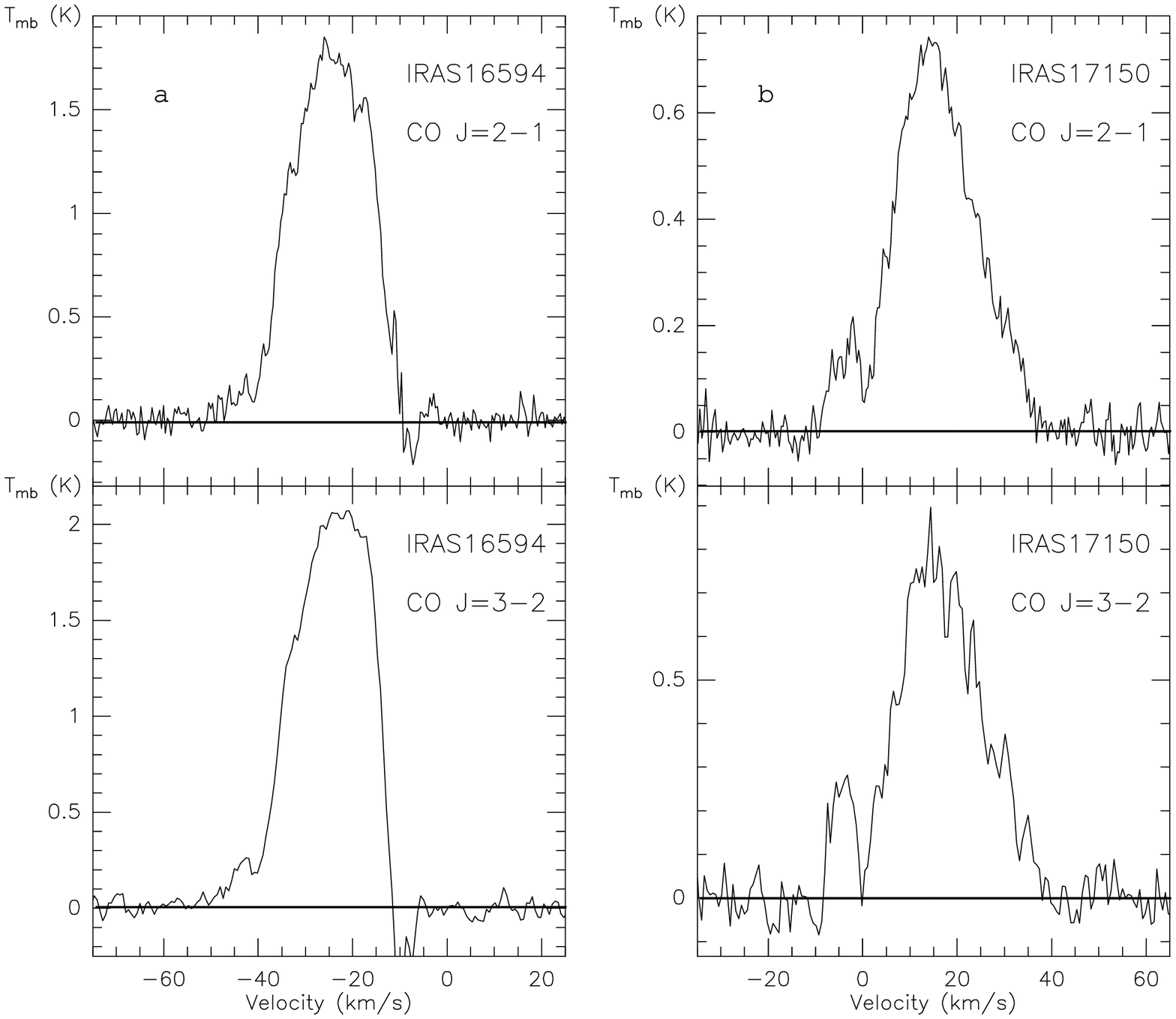}
   \caption{{\bf a) Top.} A position-switched high-resolution CO
   $J$=2-1 spectrum toward \object{IRAS16594-4656}. {\bf Bottom.} A
   beam-switched CO $J$=3-2 spectrum toward
   \object{IRAS16594-4656}. \newline {\bf b) Top.} A position-switched
   high-resolution CO $J$=2-1 spectrum toward
   \object{IRAS17150-3224}. {\bf Bottom.} A beam-switched CO $J$=3-2
   spectrum toward \object{IRAS17150-3224}.}
   \label{figCOspec}
   \end{figure*}

   \begin{figure*}
   \centering
   \includegraphics[width=9cm]{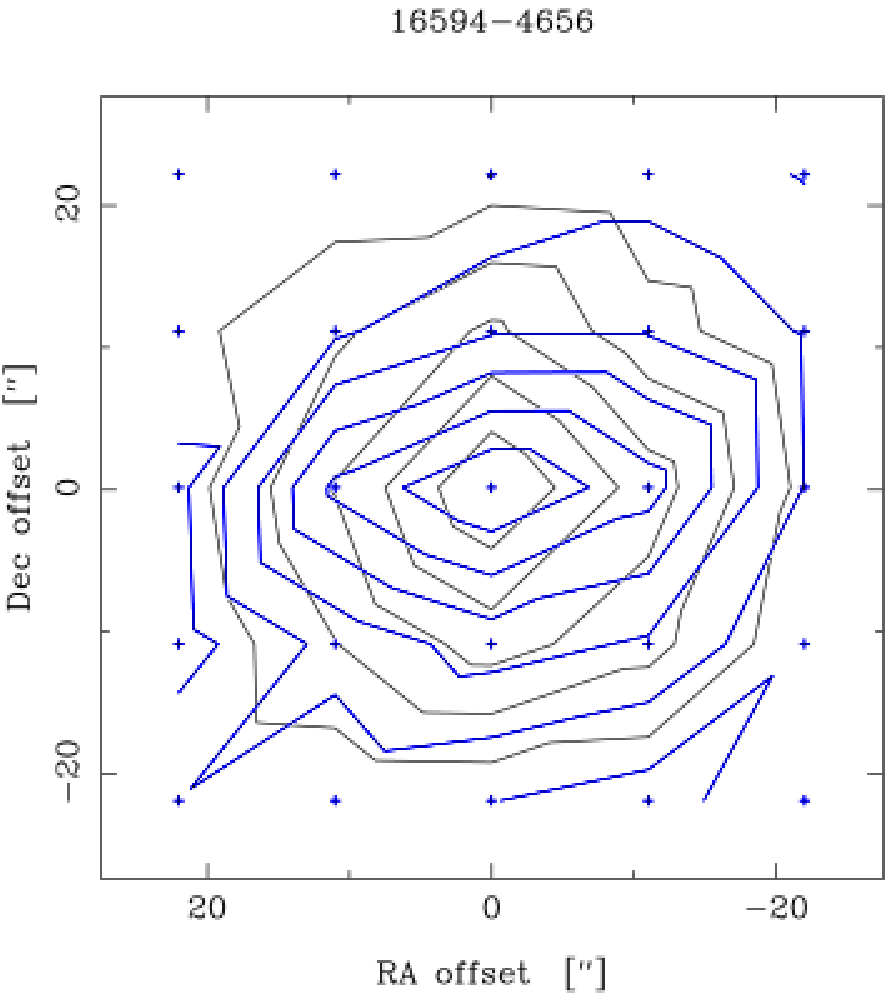}\includegraphics[width=9cm]{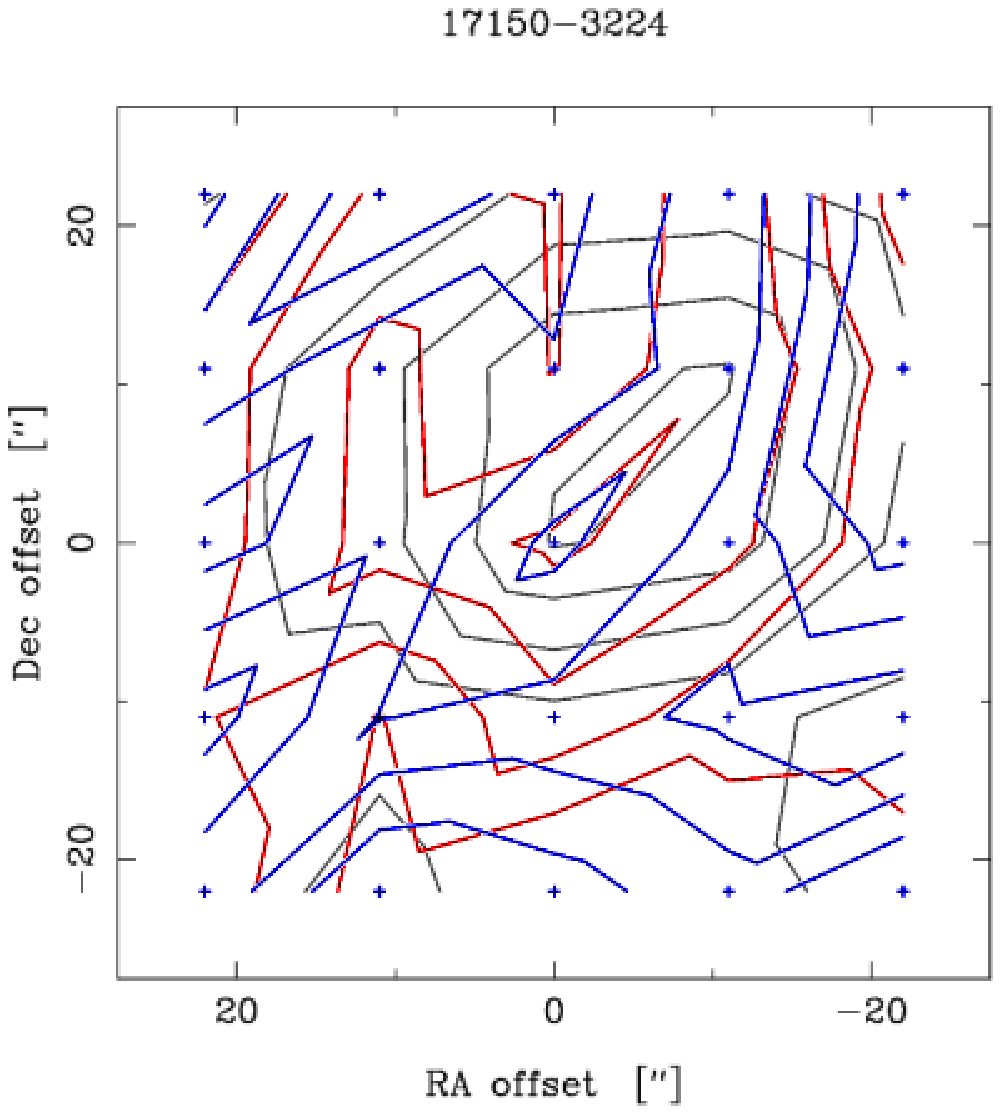}
   \caption{Position-position contour emission maps of \object{IRAS16594-4656},
   and \object{IRAS17150-3224}. Blue contours show blue-shifted emission, black
   contours emission from the centre of the line profile, and red
   contours red-shifted emission.}  \label{figRBmap}
   \end{figure*}

\subsection{\object{IRAS16594-4656}}

   \begin{figure*}
   \centering{ \includegraphics[width=14cm,angle=270]{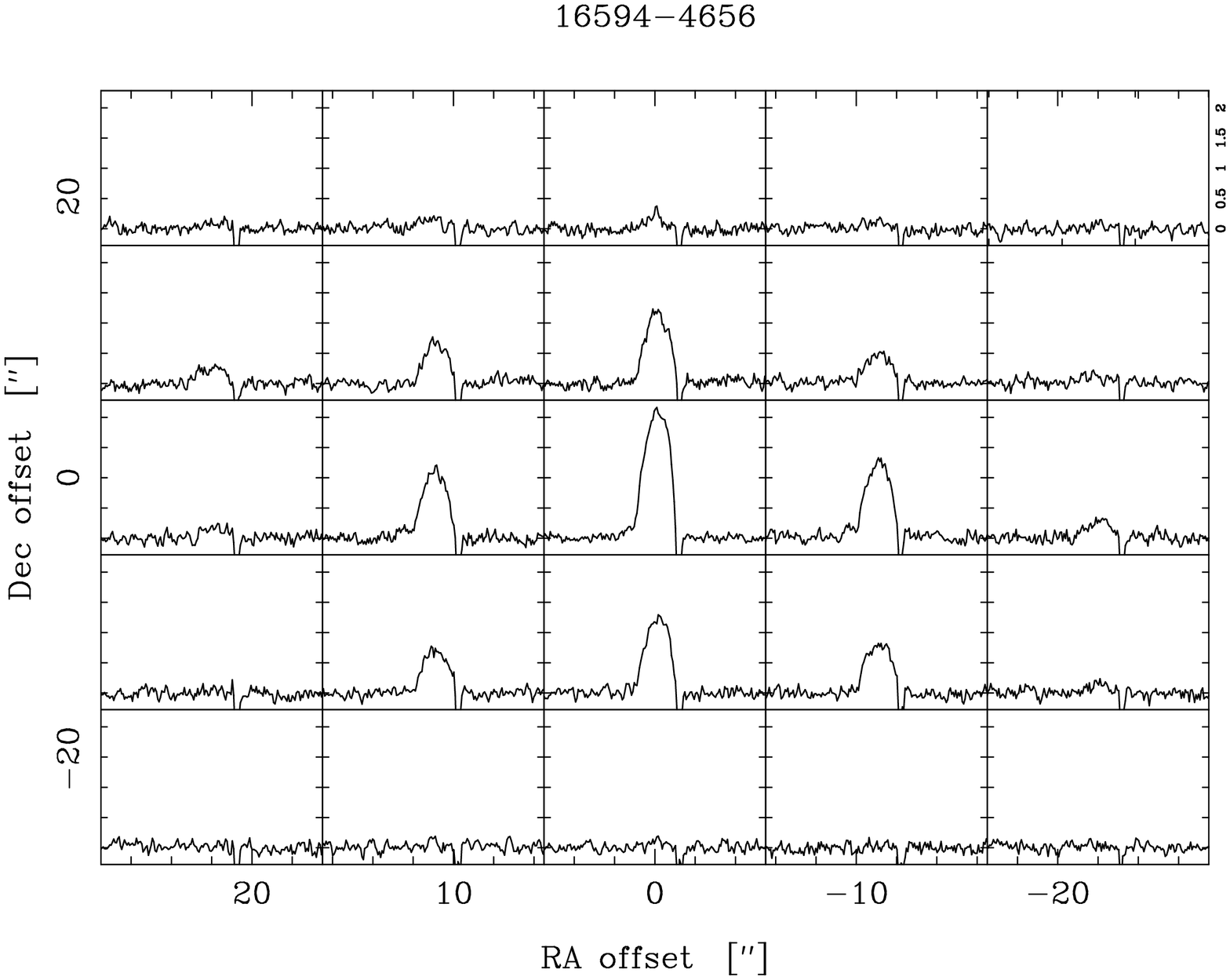}
   \caption{A map of the CO $J$=2-1 line around
   \object{IRAS16954-4656}. The beam-spacing is 11\arcsec. The spectra
   are affected by interstellar line contamination in the red-shifted
   line wing.}
   \label{fig16594map}}
   \end{figure*}

The CO ($J$=2-1) and ($J$=3-2) line profiles presented in
Fig.~\ref{figCOspec}a are affected by interstellar emission in the OFF
position (between -12 and -5\,km\,s$^{-1}$\ approximately). The CO
($J$=3-2) profile is taken in dual beamswitch mode whereas the CO
($J$=2-1) profile was taken in position switch mode with an OFF
position -30\arcsec\ away in R.A. in order to minimise the effect of
the interstellar line. Despite this, a high signal-to-noise ratio
affords some indication of structure in this source, shown by features
in the line profile. A roughly parabolic profile and a source size
comparable to the beam shows that the CO emission is probably
optically thick. An outflowing wind of 14\,km\,s$^{-1}$\ is indicated
by the line profile, and there seems to be an appreciable line wing on
the blue side, possibly indicating a second wind of up to
25\,km\,s$^{-1}$. The CO ($J$=2-1) map of this source
(Fig.~\ref{fig16594map}) shows a reasonably symmetric pattern of
emission. The red/blue-shifted emission contour plot
(Fig.\ref{figRBmap}) does not show the red-shifted emission, due to
the interstellar interference. The blue-shifted emission, taken in the
interval [-50,-35]\,km\,s$^{-1}$, lies directly on top of the centre
emission, taken in the interval [-35,-12]\,km\,s$^{-1}$, indicating
that there is little or no separation in emission regions. There is
some emission at $\pm$20\arcsec\ in R.A., and an elliptical Gaussian
fit to the integrated intensity map shows that the FWHM ellipse has a
major axis of $\sim$25\arcsec\ and a minor axis of $\sim$22.5\arcsec\
(position angle of 57\degr), very similar to the size of the SEST beam
(23\arcsec) at 230\,GHz. Hence the source is on the limit of being
resolved.

\subsection{\object{IRAS17150-3224}}

   \begin{figure*}
   \centering{
   \includegraphics[width=14cm,angle=270]{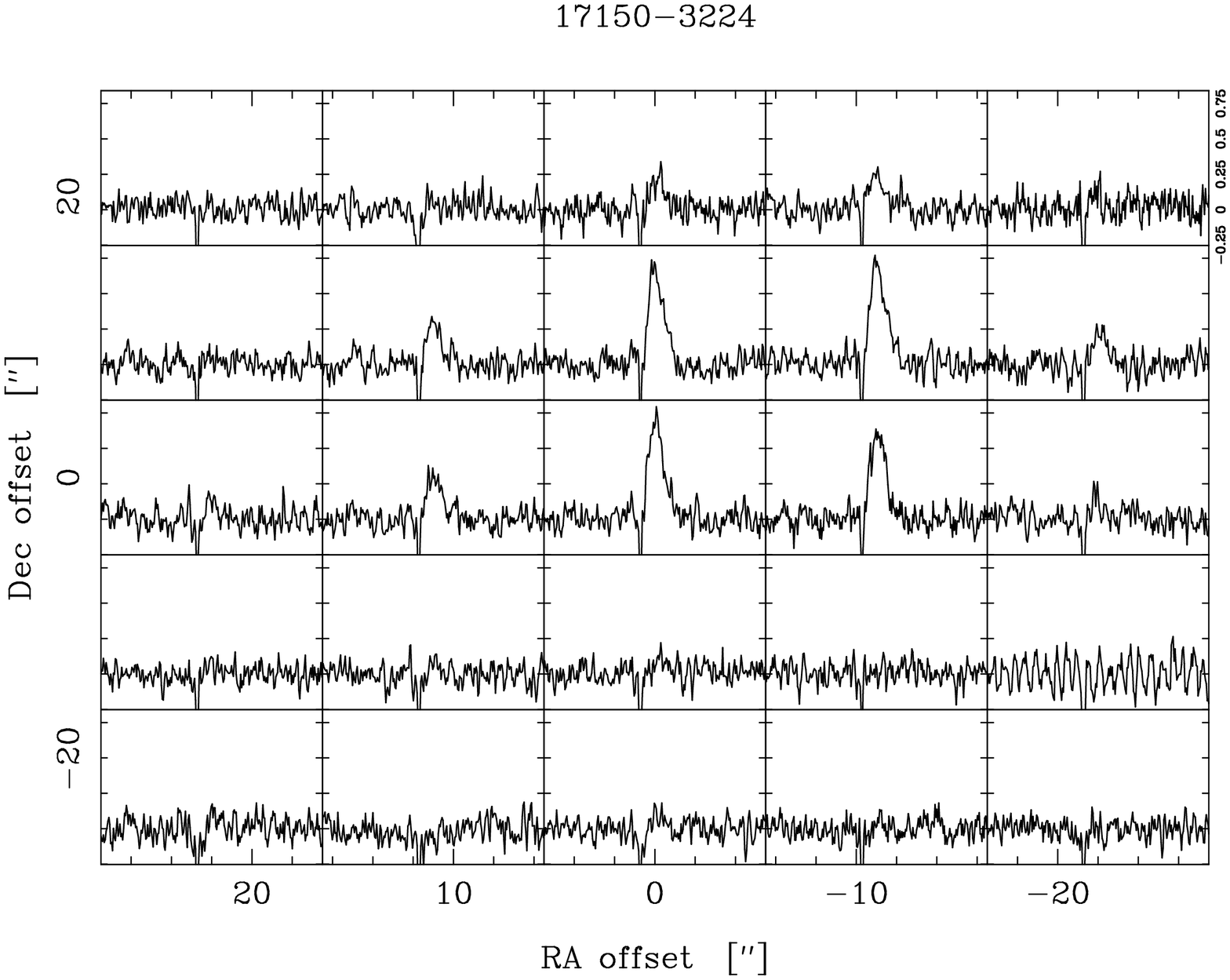}
   \caption{A map of the CO $J$=2-1 line around
   \object{IRAS17150-3224}, taken in beamswitch mode. The beam-spacing
   is 11\arcsec. The spectra are affected by interstellar line
   contamination in the blue-shifted line wing.}
   \label{fig17150map}}
   \end{figure*}

The CO ($J$=2-1) spectrum (Fig.~\ref{figCOspec}b) of
\object{IRAS17150-3224} was taken in position switch mode, with an OFF
position of +45\arcsec\ in R.A. The CO ($J$=3-2) was taken in
beamswitch mode. Both CO line profiles of \object{IRAS17150-3224}
shown in Fig.~\ref{figCOspec}b are distinctly triangular in shape,
similar to those shown by \object{89 Her}, the \object{Red Rectangle},
and \object{M2-9} (and also \object{IRAS17441-2411}, presented in this
paper). Here the triangular shape is affected by an interstellar line
in the blue-shifted wing (between 0 and 10\,km\,s$^{-1}$
approximately). The CO lines indicate a wind of some 30\,km\,s$^{-1}$\
coming from the source, and yet a parabolic fit to the line indicates
a wind of 14.5\,km\,s$^{-1}$. This latter wind is ostensibly the AGB
wind of the star, with the higher velocity emission probably coming
from an inner swept-up shell, moving at a velocity intermediate to the
AGB wind and a post-AGB or super-wind. It must be noted, however, that
a parabolic fit to a triangular lineshape will usually underestimate
the expansion velocity of the outflowing wind.

The CO ($J$=2-1) emission map (Fig.~\ref{fig17150map}) is taken in
position switch mode to minimise the effects of the interstellar
interference, and although hampered slightly by poor pointing, seems
to show a reasonably symmetric distribution which does not appear to
be resolved by the telescope beam. The contour map
(Fig.~\ref{figRBmap}) does not show any separation in the source. The
red-shifted emission was taken from the interval
[30,40]\,km\,s$^{-1}$, the centre emission from [5,30]\,km\,s$^{-1}$
and the blue-shifted emission from \mbox{[-18,-6]}\,km\,s$^{-1}$. A
two-dimensional elliptical Gaussian fit to the centre emission gives
an extent of $\sim$20\arcsec$\times$26\arcsec\ (position angle of
-160\degr) for this source. Again, this is similar to the size of the
SEST beam at 230\,GHz.

\subsection{Other sources observed in CO}

   \begin{figure*}
   \sidecaption 
   \includegraphics[width=12cm]{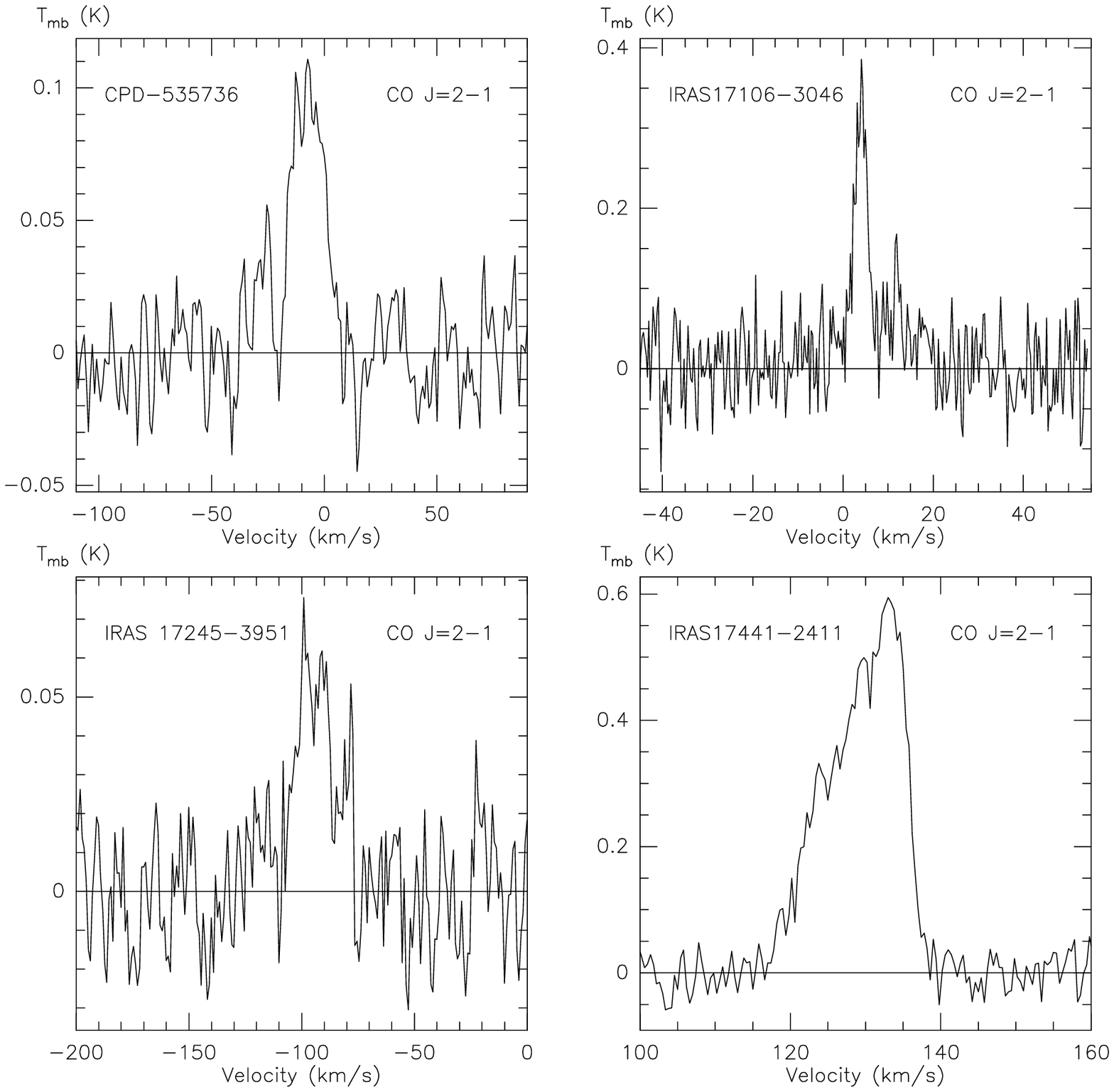}
   \caption{CO $J$=2-1 spectra of PPNe candidates, named in the upper left-hand corner of each spectrum.}
   \label{figotherco}
   \end{figure*}

Spectra of select sources which have been hitherto unobserved and/or
unpublished in the CO ($J$=2-1) line are presented in
Fig.~\ref{figotherco}. HCN ($J$=1-0) spectra have also been taken (not
presented), and HCN/CO intensity ratios are derived
(Table~\ref{iiothers}).

{\it\object{CPD-53\degr5736}}. Although not clear from the noisy
spectrum, \object{CPD-53\degr5736} seems to show the presence of a
wind expanding at 15\,km\,s$^{-1}$.

{\it\object{IRAS17106-3046}}. This object shows a strange profile:
some low-level emission, perhaps a wind of $\sim$15\,km\,s$^{-1}$, and
a thin, narrow spike. This narrow spike (of width
$\sim$2\,km\,s$^{-1}$) could correspond in some way to the
circumstellar disk observed by \citet{Kwok_etal2000}. The narrow
feature seems to be offset from the centre of the broad feature,
although the signal-to-noise ratio is low.

{\it\object{IRAS17245-3951}}. Low signal-to-noise means that this
profile is hard to qualify. The profile suggests an expansion
velocity of 15\,km\,s$^{-1}$, but this value does have a considerable
uncertainty.

{\it\object{IRAS17441-2411}}. This source has a triangular lineshape,
similar to \object{IRAS17150-3224}. It is somewhat asymmetric, and
quite narrow in velocity (20\,km\,s$^{-1}$\ in width). There are no
appreciable line wings.

\begin{table*}
   \caption{Observed lines.}
\label{iiothers}
   \begin{flushleft}
   \begin{tabular}{l cc cc cc cc cc cc}
   \hline\hline
 &\multicolumn{2}{c}{\object{CPD-53\degr5736}}&\multicolumn{2}{c}{\object{IRAS16594--4656}}&\multicolumn{2}{c}{\object{IRAS17106--3046}}&\multicolumn{2}{c}{\object{IRAS17150--3224}}&\multicolumn{2}{c}{\object{IRAS17245--3951}} &\multicolumn{2}{c}{\object{IRAS17441--2411}} \\
\cline{2-3}\cline{4-5}\cline{6-7}\cline{8-9}\cline{10-11}\cline{12-13}
Molecule & $T_{\mathrm{mb}}$ & $\int T_{\rm mb}{\rm dv}$ & $T_{\mathrm{mb}}$ & $\int T_{\rm mb}{\rm dv}$ & $T_{\mathrm{mb}}$ & $\int T_{\rm mb}{\rm dv}$ & $T_{\mathrm{mb}}$ & $\int T_{\rm mb}{\rm dv}$ & $T_{\mathrm{mb}}$ & $\int T_{\rm mb}{\rm dv}$ & $T_{\mathrm{mb}}$ & $\int T_{\rm mb}{\rm dv}$ \\ 
 & K & K\,km\,s$^{-1}$ &  K & K\,km\,s$^{-1}$ &  K & K\,km\,s$^{-1}$ &  K & K\,km\,s$^{-1}$ & K & K\,km\,s$^{-1}$ & K & K\,km\,s$^{-1}$ \\
    \hline
HCN($J$=1-0) & ---  & $<$\,0.07 & 0.02 & \phantom{0}0.47 & ---  & ---  & 0.02 & \phantom{0}0.54 & ---  & $<$\,0.09  & 0.05 & 0.32 \\
CO($J$=2-1) & 0.09 & \phantom{$<$\,}2.07 & 1.75 & 32.60 & 0.30 & 1.84 & 0.65 & 12.84 & 0.04 & \phantom{$<$\,}0.58       & 0.57 & 6.24 \\
I$_{\rm HCN}$/I$_{\rm CO}$&      & $<$\,0.04 &   & \phantom{0}0.01 &    & --- &   & \phantom{0}0.04 &     & $<$\,0.15  &      & 0.05 \\
  \hline
   \end{tabular}
   \end{flushleft}
\end{table*}

\subsubsection{I$_{\rm HCN}$/I$_{\rm CO}$\ ratios}

The I$_{\rm HCN}$/I$_{\rm CO}$\ ratios shown in Table~\ref{iiothers}
($\sim 0.01-0.05$) look to be reasonably consistent throughout
the sample of PPN candidates. To put these figures into context, a
sample of oxygen stars selected by \citet{Lindqvist_etal1988} has an
average I$_{\rm HCN}$/I$_{\rm CO}$\ ratio of 0.124, whereas the carbon
star sample of \citet{Olofsson_etal1990} has an average I$_{\rm
HCN}$/I$_{\rm CO}$\ ratio of 0.623. Both these samples were made up of
AGB stars; the values for PPNe seem to be
smaller. \object{IRAS07134+1005} and \object{IRAS19500-1709}, both
C-rich PPNe, have I$_{\rm HCN}$/I$_{\rm CO}$\ ratios of 0.15 and 0.08,
respectively \citep{Bujarrabal_etal1992}. Evolved PNe generally have
I$_{\rm HCN}$/I$_{\rm CO}$\ ratios of less than 0.1
\citep[c.f.,][]{Bachiller_etal1997b}.

\section{CO line modelling}
\label{modelling}
A detailed, non-LTE, radiative transfer code based on the Monte Carlo
method was used in order to model the observed molecular line
emission. The code assumes the emitting envelope to be spherically
symmetric and expanding at a constant velocity. The thermal balance
equation, where CO is assumed to be the main molecular coolant, is
solved together with the molecular excitation in order to obtain the
kinetic temperature structure. The CO observations are used as
constraints in the modelling to determine also the density and
velocity fields through a $\chi^2$-analysis. This approach was
also adopted by \citet{Woods_etal2003b} when modelling a sample of
high mass-loss rate carbon stars. More details on the radiative
transfer modelling can be found in \citet{Schoeier01}. The code
has been benchmarked to a high accuracy \citep{vanZadelhoff02}.

In order to derive the mass loss rate which, together with the
expansion velocity, sets the density scale in the wind through the
continuity equation, a CO abundance relative to H$_2$ must be
adopted. For \object{IRAS16594-4656} a value of $1\times10^{-3}$,
which is typical of a C-rich AGB envelope, is used, whereas for the
O-rich object \object{IRAS17150-3224} a value of $2\times10^{-4}$ is
adopted.  The derived envelope properties such as the mass loss rate
($\dot{M}$), expansion velocity (v$_\mathrm{exp}$), extent of the CO
envelope ($R_\mathrm{e}$) and $^{12}$CO/$^{13}$CO-ratio are reported
in Table~\ref{modeldata} together with the adopted distance ($D$),
luminosity ($L$) and initial CO fractional abundance
($f_{\mathrm{CO}}$). Also shown is the $h$-parameter that determines
the amount of heating in the envelope due to momentum transfer due to
dust-gas collisions \citep[see][ for details]{Schoeier01}.

\begin{figure*}
   \centering{   
   \includegraphics[width=17.2cm]{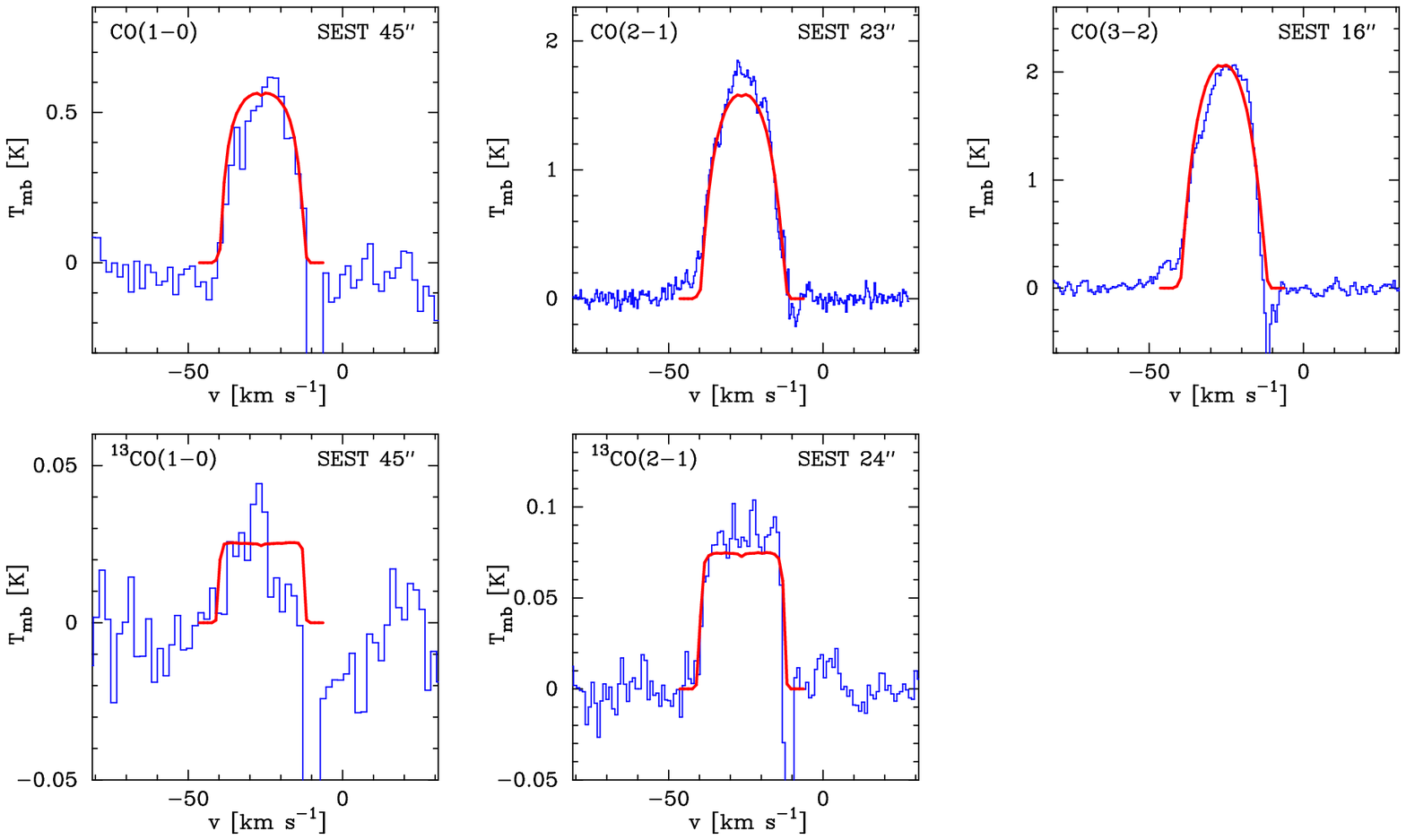}
   \caption{Best fit CO model of \object{IRAS16594-4656} (solid line) overlayed on observations (histogram) using a mass loss rate of $1\times10^{-5}$~M$_{\odot}$\,yr$^{-1}$ and a $^{12}$CO/$^{13}$CO-ratio of 30.}
   \label{fig16594_mod}}
   \end{figure*}

In Fig.~\ref{fig16594_mod} the best fit model for
\object{IRAS16594-4656}, using a mass loss rate of
$1\times10^{-5}$~M$_{\odot}$\,yr$^{-1}$\ and a
$^{12}$CO/$^{13}$CO-ratio of 30, is overlayed on the observed
spectra. The total integrated intensities in the lines are well
reproduced as are the overall line profiles. There is confusion due to
interstellar lines at the red-shifted edge of the line profiles. Also,
on the blue-shifted edge there are signs of a second weak component,
possibly a higher velocity wind due to the present-day mass loss.

\begin{figure*}
   \centering{   
   \includegraphics[width=17.2cm]{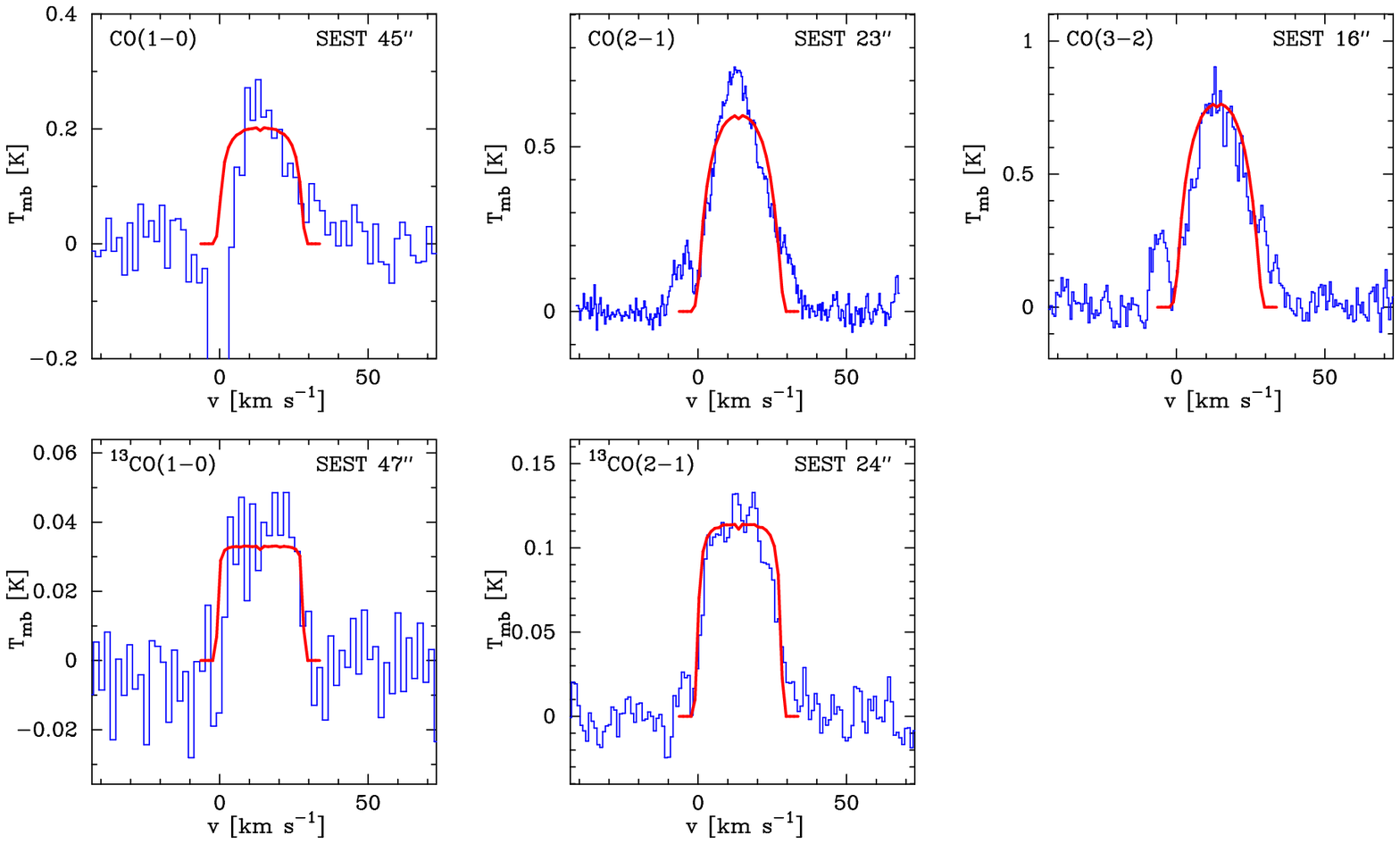}
   \caption{Best fit CO model of \object{IRAS17150-3224} (solid line) overlayed on observations (histogram) using a mass loss rate of $3\times10^{-5}$~M$_{\odot}$\,yr$^{-1}$ and a $^{12}$CO/$^{13}$CO-ratio of 7.}
   \label{fig17150_mod}}
   \end{figure*}

A good fit to CO observations is also obtained for
\object{IRAS17150-3224} using a mass loss rate of
$3\times10^{-5}$~M$_{\odot}$\,yr$^{-1}$\ (Fig.~\ref{fig17150_mod}).
The $^{12}$CO/$^{13}$CO-ratio is 7, significantly lower than for
\object{IRAS16594-4656} but consistent with \object{IRAS17150-3224}
being O-rich \citep{Abia01}. The $^{12}$CO spectra for this source
show signs of excess emission at larger expansion velocities possibly
indicating the presence of a faster moving wind. However, the blue
side of the emission is confused by interstellar line contamination.

The $h$-parameter of 10 derived for \object{IRAS16594-4656} is
significantly larger than that of \object{IRAS17150-3224} and that
which is typically derived for high mass-loss rate AGB-stars, $h\sim
1-2$, \citep{Schoeier01,Woods_etal2003b}. The kinetic temperature
derived through the envelope of \object{IRAS16594-4656} is about a
factor of two, on average, larger than that obtained for the carbon
star \object{IRC+10216} \citep{SchoeierOlofsson2000,Schoeier01}.
Such a large discrepancy is hard to explain due to, e.g.,
uncertainties in the adopted distance, and could instead indicate that
the properties of the dust grains are different in
\object{IRAS16594-4656} or that there is an additional source of
heating of the envelope. A further possibility is that the spectra
are contaminated by a second, warmer component such as the bipolar
present day mass-loss seen in optical images at smaller spatial
scales.

The winds of these PPNe are assumed to resemble those of their
progenitor AGB-stars, i.e., spherical symmetry and a constant
expansion velocity are assumed. Given that both sources left the AGB
$\sim 200-400$~yr ago (see Sect.~\ref{chemev}) this assumption
should be valid for distances $\gtrsim 1\times10^{16}$\,cm, since the
material outside this radius would once have made up the AGB
wind. However, there is the possibility that a high velocity bipolar
wind is penetrating this remnant AGB shell. The CO emission from the
lower rotational transitions observed here are formed mostly at
distances larger than $5\times10^{16}$~cm and thus mainly probe the
outer, AGB-type, wind.  For other molecular species (e.g., HCN)
contributions from the present-day wind might become significant.

However, a more detailed treatment of the wind characteristics will
require high-resolution, multi-transition observations that will
become a reality with the upcoming APEX/ALMA telescopes. Deviations
from the AGB wind model will become more apparent in the high-$J$\ CO
lines, both in the form of the line profile (broader wings) and in the
intensity. The AGB wind model predicts that the higher-$J$\ CO lines
($J$=4-3 to 13-12) should all have an integrated intensity of about
30\,K\,km\,s$^{-1}$\ and 12\,K\,km\,s$^{-1}$\ for
\object{IRAS16594-4656} and \object{IRAS17150-3224}, respectively,
when observed with the APEX telescope. Also, the effect of dust
emission in the excitation of the molecules, which is not taken into
account in the present analysis, needs to be investigated.

\begin{table*}
   \caption{Adopted and derived modelling parameters of the two PPNe.}
\label{modeldata}
   \begin{flushleft}
   \begin{tabular}{lcccccccc}
   \hline\hline
 IRAS No. & $D$     & $L$       & $f_{\mathrm{CO}}$       & $\dot{M}$               & v$_\mathrm{exp}$& $R_\mathrm{e}$      & $h$ & $^{12}$CO/$^{13}$CO \\
          & [kpc]   & [L$_{\odot}$]    & & [M$_{\odot}$\,yr$^{-1}$]& [km\,s$^{-1}$]   & [cm]                 &     &  \\
    \hline
16594-4656& 1.80$^a$& \phantom{0}4\,900& $1\times$10$^{-3}$ & $1\times$10$^{-5}$      & 14.0             & 2.9$\times$10$^{17}$& 10.0&30 \\
17150-3224& 2.42$^b$& 11\,000      & $2\times$10$^{-4}$    & 3$\times$10$^{-5}$      & 14.5             & 2.4$\times$10$^{17}$& \phantom{0}1.8& \phantom{0}7\\
  \hline
   \end{tabular}
  {\footnotesize
   \begin{enumerate}
        \renewcommand{\labelenumi}{(\alph{enumi})}
        \item \citet{VDV_etal1989}.
	\item \citet{Bujarrabal_etal2001}
   \end{enumerate}
     }
\end{flushleft}
\end{table*}

\section{Molecular evolution}

\subsection{Molecules}

The molecules seen in PPNe give information on various processes which
have occurred in the history of the particular PPN. HCN and CN can be
used to give an idea of the evolutionary status of a PPN: in AGB
stars, HCN is usually more abundant than CN. As HCN becomes
increasingly photodissociated, its daughter, CN, becomes increasingly
abundant; i.e., one would expect the CN/HCN ratio to increase with
evolutionary age, in the early post-AGB phase (see
Fig.~\ref{modelres}). A similar effect is seen in the comparison of
HCN and HNC: In \object{CRL618}, for example, the column densities of
HCN and HNC are comparable \citep{HerpinCern2000}, whereas in
\object{IRC+10216} (a less evolved object) HCN is far more abundant
that HNC. In both these objects HNC is formed via dissociative
recombination of HCNH$^+$.

The HCN/CO ratio should also give some indication of age, 
since photodissociation would cause this ratio to decrease with time.
However, it may be difficult to compare a number of objects
which includes a mixture of O-rich and C-rich stars, since they may
have had a different HCN/CO ratio when they started to move off the
AGB.

The presence of CS can be used in a similar way. CS is abundant in AGB
envelopes, and although it can be enhanced by shocks in PPNe
\citep{Kasuga_etal1997,Mitchell1984} is generally of lower abundance,
and unobserved in PNe until recently \citep{Woods_etal2003c}. HC$_3$N,
SiO and SiC$_2$ demonstrate the same sort of behaviour, decreasing in
abundance going into the PPNe phase.

HCO$^+$\ becomes greatly enhanced in the PPNe phase, and yet it is
very difficult to explain why. Theoretical models of post-AGB objects
all fail to produce enough HCO$^+$\ by at least an order of magnitude,
if not more
\citep[e.g.,][]{Howe_etal1994,Ali_etal2001,Woods_etal2003a}. HCO$^+$\
is only efficiently formed by ionising radiation, and yet the levels
of cosmic ray ionising radiation required to produce the desired
levels of HCO$^+$\ result in the destruction of most other
molecules. This is due to the fact that cosmic rays induce UV photons
in their interaction with H$_2$; these photons dissociate and ionise
molecules. \citet{Woods_etal2003a} considered whether X-rays might be
the responsible for the high level of HCO$^+$. Their treatment was
quite general -- a treatment where X-rays are solely responsible for
ionisation (and not dissociation) might prove interesting. This
approach was taken in the chemical model explained in
Sect.~\ref{chemmodel}. However, a more thorough treatment needs
to be made.

HCO$^+$\ can also form in photon-dominated regions (PDRs), in warm gas
(800\,K), through reactions involving CO$^+$, which itself is formed
through reactions between C$^+$\ and OH
\citep{Hasegawa_etal2000}. Moreover, shocked gas can be identified by
the presence of HCO$^+$\ (as well as SiO)
\citep[e.g.,][]{MitchellWatt1985}, and although high levels of
HCO$^+$\ do not necessarily imply shocks, the presence of high levels
of SiO does give a much better indication of the presence of shocked
gas.

The $^{12}$CO/$^{13}$CO is low ($<10$) in several PPNe
\citep[e.g.,][]{Kahane_etal1992} compared to \mbox{$10<$
$^{12}$CO/$^{13}$CO $<25$} in evolved PNe
\citep{Palla_etal2000,Bachiller_etal1997b}, \mbox{$10<$
$^{12}$CO/$^{13}$CO $<40$} in compact PNe
\citep{JosselinBachiller2003} and \mbox{$10\lesssim$
$^{12}$CO/$^{13}$CO $\lesssim50$} in AGB stars
\citep[e.g.,][]{Kahane_etal1988, GreavesHolland1997,
SchoeierOlofsson2000}.  From our radiative transfer analysis we derive
$^{12}$CO/$^{13}$CO ratios of 30 and 7 for the PPNe
\object{IRAS16594-4656} and \object{IRAS17150-3224}, respectively.
This ratio may be an indicator of evolution (since $^{13}$CO
self-shields less than $^{12}$CO), but it is more likely to be an
indicator of a star's nucleosynthesis history which may be very
different for otherwise apparently similar objects
\citep[e.g.,][]{Kahane_etal1992,Kahane_etal1988}.

\subsection{Fractional abundances}

\subsubsection{Calculation of fractional abundances}
\label{abunds}

In Sect.~\ref{modelling} the mass loss rates, and hence the radial
distribution of the number density of H$_2$, for the sample PPNe were
determined from a detailed radiative transfer analysis of the observed
CO line emission adopting a particular CO/H$_2$ abundance ratio. Here,
fractional abundances for the remaining molecular species observed
were calculated assuming optically thin emission. The radiative
transfer for these lines is treated in the same way as described in
\citet{Woods_etal2003b}, and the results are shown in
Table~\ref{fracabunds}. A constant excitation temperature of 25\,K is
assumed throughout the emitting region.  Upper limits to fractional
abundances are given in Table~\ref{upperlimits}, and are calculated
using the integrated intensity from Eq.~\ref{uleqn}. The same
molecular data as used in \citet{Woods_etal2003b} are adopted here.

Inner and outer radii of molecular distributions are determined in
much the same way as \citet{Woods_etal2003b}, using a
photodissociation model. Here, however, there is not the benefit of
good interferometric data. Hence simplifying assumptions have to be
made about certain molecular distributions. The distribution of all
parent species, with the exception of CO, is calculated from the
photodissociation model (observed parent species are HCN, CS, SiO,
SiS). The distribution of CO comes from the radiative transfer
modelling, as described in Sect.~\ref{modelling}. Self-shielding of CO
is taken into account, in the manner of \citet{Mamon_etal1988}. All
ionisation products or products of the circumstellar chemistry are
assumed to take on the distribution of C$_2$H, with the exception of
CN, which has a slightly more expansive distribution. This should not
have too great an effect on results since there is only a linear
dependence on the difference between inner and outer radius. These are
very straightforward assumptions, which ignore effects of
circumstellar chemistry, and freeze-out onto grains, for
example. Again, isotopomers are assumed to have the same distribution
as their more abundant forms.

\begin{table*}
   \caption{Calculated fractional abundances, with respect to $n$(H$_2$).}
\label{fracabunds}
   \begin{flushleft}
   \begin{tabular}{lll cccccc cccccc}
   \hline\hline
             &&                     &&           & \object{IRAS16594--4656} &                 &&           &\object{IRAS17150--3224} & \\
\cline{5-7}    \cline{9-11}
Molecule     && Transition          &&$r_{\rm i}$&$r_{\rm e}$& $f_{\rm X}$            &&$r_{\rm i}$&$r_{\rm e}$& $f_{\rm X}$\\
             &&                     && cm        & cm        &              && cm        & cm        &  \\
    \hline
$^{13}$CO$^a$&&	---	            && ---       & 2.9\,(17) &  3.3\,(-5)        && ---       & 2.4\,(17) &2.9\,(-5)\\
HCN	     &&	$J$=1-0	            && ---       & 2.3\,(16) &1.0\,(-6)&& ---       & 4.4\,(16) &3.8\,(-7)\\
HCN	     &&	$J$=3-2	            && ---       & 2.3\,(16) &1.9\,(-7)&& ---       & 4.4\,(16) &---\\
CN	     &&	$N$=1-0	            && 1.7\,(16) & 5.1\,(16) &9.6\,(-6)&& 3.9\,(16) & 9.8\,(16) &U.L.$^b$\\
CN	     &&	$N$=2-1	            && 1.7\,(16) & 5.1\,(16) &2.5\,(-6)&& 3.9\,(16) & 9.8\,(16) &U.L.$^b$\\
  \hline
   \end{tabular}
  {\footnotesize
   \begin{enumerate}
        \renewcommand{\labelenumi}{(\alph{enumi})}
        \item $^{13}$CO fractional abundances derived from full radiative transfer (Sect.~\ref{modelling}).
        \item signifies a non-detection, but the calculation of an upper limit to the fractional abundance has been made (see Table~\ref{upperlimits}).
   \end{enumerate}
     }
\end{flushleft}
\end{table*}

\begin{table}
   \caption{Calculated upper limits to fractional abundances, with respect to $n$(H$_2$).}
\label{upperlimits}
   \begin{flushleft}
   \begin{tabular}{ll ccc}
   \hline\hline
             && \object{IRAS16594--4656} && \object{IRAS17150--3224} \\
\cline{3-3}    \cline{5-5}
Molecule     && $f_{\rm X}$     && $f_{\rm X}$\\
    \hline
CO$^+$	     && 1.1\,(-8)      &&	---	        \\
HCO$^+$	     &&	1.2\,(-7)      &&	---	        \\
H$_2$CO	     &&	1.1\,(-6)      &&	---	        \\
CS	     &&	6.5\,(-8)      &&	---	        \\
C$_2$S	     &&	3.4\,(-7)      &&	9.5\,(-8) 	\\
C$_3$S	     &&	4.1\,(-7)      &&	---     	\\
SiO	     &&	5.1\,(-8)      &&	---	        \\
SiS	     &&	3.2\,(-7)      &&	---	        \\
SiC$_2$	     &&	4.7\,(-7)      &&	2.1\,(-6)       \\
CN	     &&	Det.$^a$       &&	1.1\,(-8) 	\\
$^{13}$CN    &&	5.0\,(-7)      &&	---	        \\
H$^{13}$CN   &&	2.4\,(-8)      &&	1.3\,(-8) 	\\
HNC	     &&	1.8\,(-7)      &&	---	        \\
HN$^{13}$C   &&	1.9\,(-7)      &&	---	        \\
HC$_3$N	     &&	2.7\,(-7)      &&	---	        \\
H$^{(13)}$C$_3$N$^b$  &&	2.7\,(-7)      &&	---	        \\
H$^{13}$CCCN &&	2.0\,(-7)      &&	---	        \\
C$_3$N	     &&	4.2\,(-7)      &&	---	        \\
CH$_3$CN     &&	8.2\,(-8)      &&	5.3\,(-8)       \\
C$_2$H	     &&	3.1\,(-6)      &&	---	        \\
CH$_3$C$_2$H &&	3.5\,(-6)      &&	---	        \\
C$_3$H$_2$   &&	2.5\,(-7)      &&	---	        \\
C$_4$H	     && 4.5\,(-6)      &&	1.4\,(-6)       \\
  \hline
   \end{tabular}
  {\footnotesize
   \begin{enumerate}
        \renewcommand{\labelenumi}{(\alph{enumi})}
	\item signifies a detection has been made of that
	particular line in that particular source (see
	Table~\ref{fracabunds}).
	\item signifies a blend of HCC$^{13}$CN and HC$^{13}$CCN.
   \end{enumerate}
     }
\end{flushleft}
\end{table}

\begin{table*}
   \caption{Comparison with fractional abundances in other PPNe, arranged in order of decreasing stellar temperature.}
\label{litabunds}
 \little
   \begin{flushleft}
   \begin{tabular}{l ccccccccccc}
   \hline\hline
C-rich PPNe:& \object{IRAS16594-4656}     && \object{M1-16}              && \object{CRL618}                       && \object{CRL2688}                                    && \object{IRAS19500-1709} && \object{IRC+10216} \\
\hline
$^{13}$CO & \phantom{$<$\,}3.3\,(-5)$^{a}$  && && 2.4\,(-5)$^b$                && 5.2\,(-5)$^b$         && 2.0\,(-5)$^k$ &&        \\
HCN       & \phantom{$<$\,}6.1\,(-7)$^a$      && \phantom{$<$\,}1.0\,(-7)$^j$      && 4.4\,(-6)$^b$--5.0\,(-7)$^d$ && \phantom{$<$\,}2.1\,(-5)$^b$--4.0\,(-7)$^d$ && 1.5\,(-7)$^k$ && 1.3\,(-5)$^r$\\
CN        & \phantom{$<$\,}6.0\,(-6)$^a$      && \phantom{$<$\,}8.5\,(-7)$^j$      && 2.1\,(-6)$^c$                && 1.0\,(-6)$^c$                               && && 2.2\,(-6)$^r$ \\
H$^{13}$CN& $<$\,2.4\,(-8)$^a$ && && 6.3\,(-8)$^e$                && $>$\,4.0\,(-6)$^g$--5.1\,(-7)$^e$           && && 2.8\,(-7)$^r$ \\
HNC       & $<$\,1.8\,(-7)$^a$ && && 1.9\,(-6)$^b$--1.9\,(-7)$^e$ && \phantom{$<$\,}1.8\,(-7)$^h$--5.0\,(-8)$^g$ && && 5.5\,(-8)$^r$ \\
HC$_3$N   & $<$\,2.7\,(-7)$^a$ && && 4.4\,(-7)$^e$--1.9\,(-7)$^b$ && 1.5\,(-7)$^{e,g}$                           && && 1.0\,(-6)$^r$\\
HCO$^+$   & $<$\,1.2\,(-7)$^a$ && \phantom{$<$\,}9.3\,(-8)$^j$   && 2.0\,(-7)$^d$                &&                       &&                     && \\
CS        & $<$\,6.5\,(-8)$^a$ && $<$\,8.8\,(-8)$^j$ && 4.1\,(-7)$^b$--6.0\,(-8)$^d$ && 2.1\,(-6)$^b$         && && 8.8\,(-7)$^r$        \\
C$_2$H    & $<$\,3.1\,(-6)$^a$ && && 2.0\,(-6)$^d$                && 9.3\,(-6)$^f$         && && 2.6\,(-6)$^r$\\
C$_4$H    & $<$\,4.5\,(-5)$^a$ && && 2.9\,(-6)$^b$--8.0\,(-8)$^d$ &&                       && && 3.2\,(-6)$^r$\\
SiO       & $<$\,5.1\,(-8)$^a$ && $<$\,2.4\,(-8)$^j$&& $<$\,5.5\,(-7)$^b$           &&                       && && 1.1\,(-7)$^r$        \\
SiS       & $<$\,3.2\,(-7)$^a$ && $<$\,4.2\,(-8)$^j$&&                              && 4.0\,(-8)$^g$         && && 9.5\,(-7)$^r$        \\
\hline\hline
O-rich PPNe:& \object{IRAS17150-3224} && \object{M1-92} && \object{HD101584} && \object{IRAS19114+0002} && \object{OH17.7-2.0} && \object{OH231.8+4.2}          \\
\hline
$^{13}$CO & \phantom{$<$\,}2.9\,(-5)$^{a}$ && && \phantom{$<$\,}5.0\,(-4)$^q$ && \phantom{$<$\,}3.8\,(-4)$^b$ && \phantom{$<$\,}4.2\,(-6)$^b$ && \phantom{$<$\,}1.0\,(-4)$^l$--2.2\,(-5)$^b$ \\
HCN       & \phantom{$<$\,}3.8\,(-7)$^a$ && $<$\,2.0\,(-7)$^{o,p}$ && $<$\,1.7\,(-6)$^q$ && $<$\,4.5\,(-6)$^b$ && $<$\,1.6\,(-7)$^b$ && \phantom{$<$\,}4.4\,(-6)$^b$--5.2\,(-8)$^m$ \\
CN        & $<$\,1.1\,(-8)$^a$ && && && && && $<$\,2.0\,(-7)$^{c,l}$ \\
H$^{13}$CN& $<$\,1.3\,(-8)$^a$ && && && && && \phantom{$<$\,}4.8\,(-8)$^l$ \\
C$_4$H    & $<$\,1.4\,(-6)$^a$ && $<$\,3.6\,(-7)$^q$ && && $<$\,1.4\,(-6)$^b$ && $<$\,4.7\,(-8)$^b$ && \\
  \hline
   \end{tabular}
  {\footnotesize
   \begin{enumerate}
        \renewcommand{\labelenumi}{(\alph{enumi})}
	  \ \newline
	  (a) This work.
	  (b) \citet{Bujarrabal_etal1994} 
          (c) \citet{Bachiller_etal1997a}
	  (d) \citet{Bujarrabal_etal1988}
	  (e) \citet{Sopka_etal1989}
	  (f) \citet{Fuente_etal1998}
	  (g) \citet{NguyenBieging1990}
	  (h) \citet{Kasuga_etal1997}
	  (j) \citet{Sahai_etal1994}
	  (k) Using the results of \citet{Bujarrabal_etal1992,Bujarrabal_etal2001}.
	  (l) \citet{Morris_etal1987}
	  (m) \citet{Sanchez_etal1997}
	  (n) \citet{Omont_etal1993}
	  (o) \citet{Lindqvist_etal1992}
	  (p) \citet{Nercessian_etal1989} 
	  (q) Using the results of \citet{OlofssonNyman1999}.
	  (r) \citet{Woods_etal2003b}
   \end{enumerate}
     }
\end{flushleft}
\end{table*}

\subsubsection{Uncertainties}
\label{errors}

As discussed in \citet{Woods_etal2003b}, there are inherent
uncertainties in the approach taken here. Typical distance estimates
can vary by up to a factor 2, and this influences the mass-loss rate
derived from the radiative transfer modelling (see
Sect.~\ref{modelling}). When combined with the uncertainties in the
choice of inner and outer radii for the molecular distributions, the
overall error in fractional abundance varies with $\sim$$D^{-1-0}$,
where $D$\ is the distance. Given also that there are likely to be
errors introduced by the choice of excitation temperature, an overall
error of a factor 5 or so is to be expected in fractional abundance
estimates. Moreover, some species other than CO are expected to be
optically thick, most notably HCN, and the optically thin
approximation would give systematically too low fractional abundances
in those cases.  As mentioned in Sect.~\ref{modelling}, contribution
to the line intensities from the present-day, non-spherically
symmetric, high-velocity, mass loss might become significant for some
molecules. A full treatment of this problem, including the effects of
radiative excitation due to dust emission, will require high spatial
resolution observations of a large number of molecules.  With these
caveats, molecular fractional abundances are believed to be
order of magnitude estimates.

\subsection{The chemical model}
\label{chemmodel}

To investigate the changes in fractional abundance of species
during the late AGB and PPN phases, a chemical model of the
circumstellar envelope was used. Results of this model are shown in
Fig.~\ref{modelres}. The chemical model of \object{IRC+10216}
constructed by \citet[][ MHB]{Millar_etal2000} has been adapted to
investigate the chemistry, particularly the photochemistry, as an AGB
star finishes its phase of mass loss, and moves into the PPN
phase. Several simplifying assumptions are made, and the MHB model has
been changed to include X-rays, and a minimal X-ray chemistry.

\subsubsection{Model parameters and simplifying assumptions}

Several parcels of gas and dust in the circumstellar envelope (CSE) of
an AGB star are assumed to flow outward at a uniform velocity of
14\,km\,s$^{-1}$, from initial radial distances of 1, 4 and
7$\times$10$^{15}$, 1, 4 and 7$\times$10$^{16}$\ and 1 and 4
$\times$10$^{17}$\,cm from the central star. A mass-loss rate of
1$\times$10$^{-5}$\,M$\odot$\,yr$^{-1}$\ is chosen, in accordance with
the molecular line modelling in Table~\ref{modeldata}. A spherical
geometry implies that the density of the homogeneous outflow drops as
$r^{-2}$. Mass-loss is assumed to end with the start of the model.

{\it Envelope heating.} During the expansion, the
\textquotedblleft kinetic temperature\textquotedblright\ of the
parcels varies according to $T$=$T_0$($r_0/r$)$^{0.79}$, where $T_0$\
increases from 100\,K to 1\,000\,K over the first 1\,000\,yr of the
model (chosen to represent the PPN phase). This heating is due to the
increasing temperature of the central star, as it evolves towards a
white dwarf.

{\it Stellar UV field.} Following the increase in stellar
temperature, the intensity of the stellar UV field also increases. An
ionisation rate similar to that of the interstellar UV field is
assumed at the start of the model. At the end of the first 1\,000\,yr
this value has increased by a factor of 2$\times$10$^5$\
\citep[c.f.,][]{HerpinCern2000}.

{\it X-rays.} X-rays from the central star or companion
\citep{Woods_etal2002} are treated in a very simple way in this
model. The X-ray extinction law of \citet{Deguchi_etal1990} is used
(see their Eq.~B3), and an X-ray energy of 1.2\,keV is assumed
($\lambda\sim$1\,nm). X-rays only play an ionising role in the
chemical reaction network, and these reactions are given the same rate
coefficients as the equivalent cosmic-ray reactions. A very low
ionisation rate of 10$^{-17}$\,s$^{-1}$\ is chosen initially; this has
risen to a rate of 10$^{-13}$\,s$^{-1}$\ by the end of the model run.

\subsubsection{Chemical reaction network}
\label{initabunds}

The reaction network is the same as that used in
\citet{Woods_etal2003a}, with the addition of a small number of X-ray
ionisation reactions. At present the network includes 3897 reactions
among 407 species in 6 elements. Initial fractional abundances are
determined via the use of an AGB steady-state model, with a mass-loss
rate of 1$\times$10$^{-5}$\,M$_\odot$\,yr$^{-1}$\ and expansion
velocity of 14\,km\,s$^{-1}$. For example, a parcel starting at a
radius of 1$\times$10$^{16}$\,cm in the model presented here will have
initial abundances which are the same as a parcel of \textquotedblleft
AGB\textquotedblright\ material at the same radial distance. The
initial abundances of parent molecules for the AGB model are chosen to
represent a carbon-rich environment, and values are taken from MHB.

\section{Discussion}
\label{disc}

\subsection{Comparison of observations with chemical model}

   \begin{figure*}
   \sidecaption  
   \includegraphics[width=12cm]{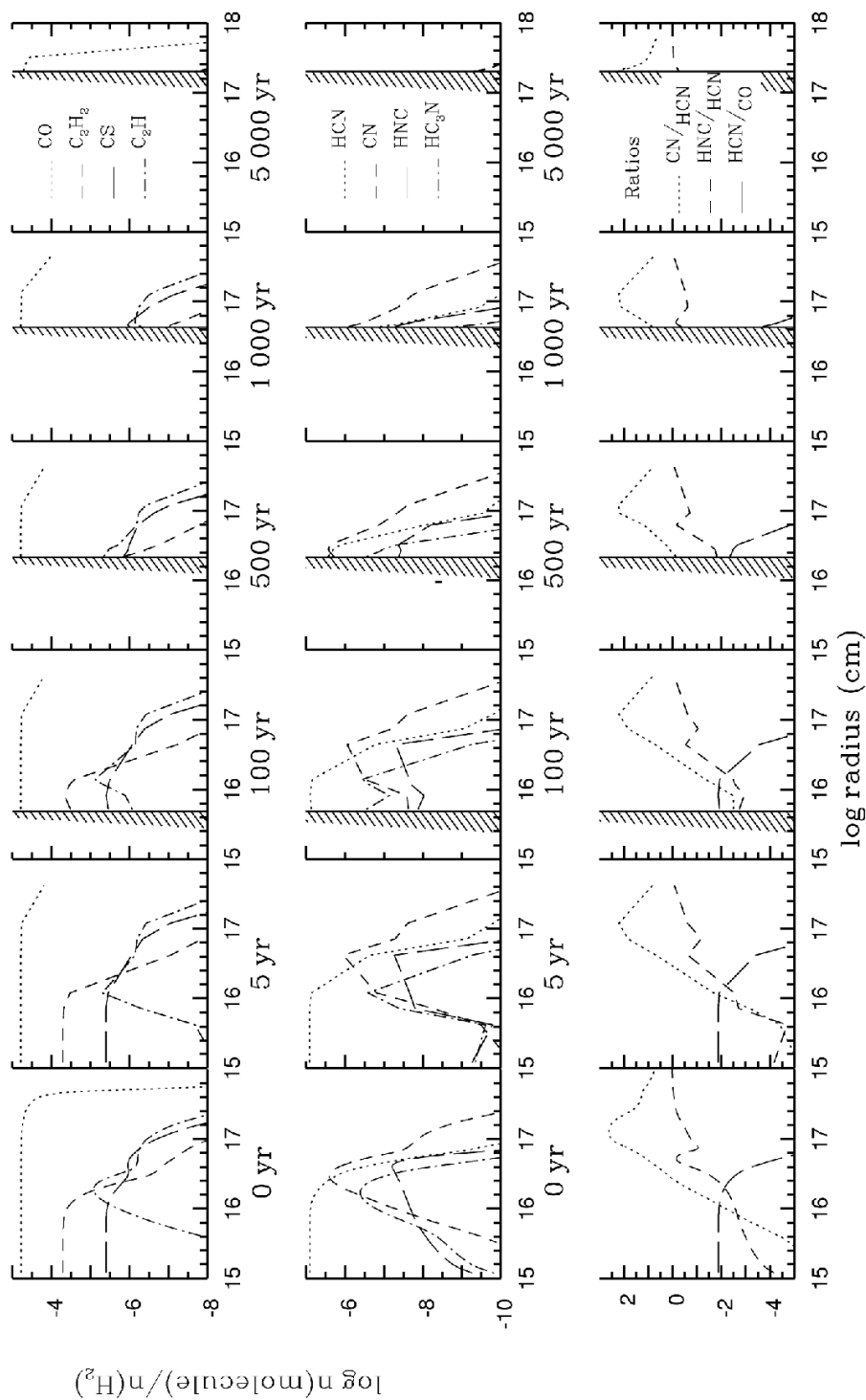}
   \caption{Results of the chemical model of
   \object{IRAS16594-4656}. The figure shows the variation of the
   fractional abundances (or ratios, in the lower row of axes) of
   selected species radially through the expanding circumstellar
   envelope at different points in time (measured in years since the
   cessation of mass-loss). The shading marks the passage of the inner
   edge of the circumstellar envelope. As can be seen at long times
   (5\,000\,yr), only CO is reasonably abundant. The three sets of
   axes labelled ``0\,yr'' show the results of a standard AGB model
   (see~\ref{initabunds}).
   \label{modelres}}
   \end{figure*}

The chemical model generally shows good agreement with the fractional
abundances derived here. Given the order of magnitude error margins
(Sect.~\ref{errors}), both HCN and CN reach the calculated fractional
abundances for \object{IRAS16594-4656} during the first 1\,000\,yr of
the model (i.e., the PPN phase). Upper limits for other observed
species (e.g., HNC, HC$_3$N, C$_2$H and C$_4$H) are also in agreement
with the model in this region. CS is predicted to be more abundant
than observed. The model also shows that a few species (e.g., CO,
C$_2$H and CN) survive at observable levels after, say, 1\,000\,yr of
post-AGB evolution (Fig.~\ref{modelres}).

Perhaps the most interesting result of the model is the variation of
the CN/HCN, HNC/HCN and HCN/CO ratios. As expected, the CN/HCN ratio
is seen to rise during the early post-AGB evolution (taking a mean
through the envelope, Fig.~\ref{modelres}). Similarly, the HNC/HCN
ratio also increases. The HCN/CO ratio declines significantly with
post-AGB age. When these three behaviours are combined, a reasonably
accurate age can be given. This approach is taken later, in
Sect.~\ref{chemev}.

\subsection{Comparison of observations with other observations}

\object{IRAS16594-4656} is very under-abundant in molecules in
comparison to C-rich AGB stars. Generally, fractional abundances
derived here, and presented in Table~\ref{fracabunds} (and the upper
limits shown in Table~\ref{upperlimits}), are an order of magnitude
deficient when compared to C-rich AGB stars, such as those surveyed in
\citet{Bujarrabal_etal1994} and \citet{Woods_etal2003b}. Two
exceptions to this are CN and $^{13}$CO, which show good agreement
with the fractional abundances derived for \object{IRC+10216}.

The lack of detection of SiO would seem to confirm that
\object{IRAS16954-4656} is C-rich, and that any silicate grains
present \citep[see ][]{Garcia_etal1999} are not being destroyed by
shocks because either shocks might not be present, or the shocks which
are present are not strong enough
\citep[c.f.,][]{VDSteeneVHoof2003}. Another explanation may be that
the shocked regions are very small, in which case high abundances of
SiO would be localised, and SiO emission diluted by the beam.

Conversely, \object{IRAS17150-3224} seems (chemically) very much like
an O-rich AGB star, such as those observed by
\citet{Bujarrabal_etal1994}, or \citet{Lindqvist_etal1988}, for
example. Although the fractional abundances of only two molecules,
$^{13}$CO and HCN, in \object{IRAS17150-3224} are suitable for
comparison, they are very similar to the O-rich AGB stars in both
AGB-star samples.

So, \object{IRAS16594-4656} and \object{IRAS17150-3224} seem to have
different chemical properties from each other, and both seem far from
the well-known molecule-rich PPNe, like \object{CRL618},
\object{CRL2688} and \object{OH231.8+4.2}. \object{IRAS16594-4656},
supposedly a C-rich object, is poor in molecules and, as can be seen
from Table~\ref{litabunds}, is at least an order of magnitude less
abundant than \object{CRL618} and \object{CRL2688} in many
molecules. However, \object{IRAS16594-4656} has generally higher
fractional abundances than \object{M1-16}, which is widely accepted to
be a standard PN. An exception to this is CN - a fractional abundance
of CN similar to those in \object{CRL618} and \object{CRL2688} may
indicate that \object{IRAS16594-4656} is a somewhat evolved PPN, but
not quite part of the PN regime yet. The theoretical work of
\citet{Woods_etal2003a} shows that fractional abundances of CN can
remain high for a period of $\sim$100 years in the PPNe phase, whilst
other molecules are destroyed, if molecular material is to be found in
the shielded environment of a circumstellar torus.

\citet{Bachiller_etal1997b} present a homogeneous sample of molecular
abundances in post-AGB objects; however, both \object{IRAS16594-4656}
and \object{IRAS17150-3224} are too low in molecular abundance to
present much of a comparison. These two PPNe are closest in fractional
abundances to the younger objects in the \citet{Bachiller_etal1997b}
sample.

Using the fractional abundance of HCN derived from the $J$=1-0 line,
since this line is least likely of the two observed to be optically
thick, a CN/HCN ratio of $<10$\ is derived for
\object{IRAS16594-4656}. This value is quite uncertain, and dependent
very much on the factors discussed in Sect.~\ref{errors}. A full
radiative transfer treatment of the HCN($J$=1-0) line does in fact
predict a fractional abundance $2-3$ times larger than that derived
via the optically thin assumption (Sect.~\ref{errors}), implying that
\object{IRAS16594-4656} could have a CN/HCN ratio as low as
$\sim3-5$. In this case the CN/HCN ratio seems to indicate that
\object{IRAS16594-4656} is just starting the post-AGB phase. Other
PPNe have a CN/HCN ratio up to an order of magnitude smaller, e.g.,
for \object{CRL618}, CN/HCN$\sim$0.5 and for \object{CRL2688},
CN/HCN$\sim$0.2. \object{IRC+10216} has a smaller CN/HCN ratio yet,
and the more evolved object \object{NGC7027} has a CN/HCN ratio of
about 10.

Both \object{IRAS16594-4656} and \object{IRAS17150-3224} have very low
HCN/CO ratios compared to \object{CRL618}, \object{CRL2688} and
\object{OH231.8+4.2}. \object{IRAS16594-4656} has an HCN/CO ratio of
6$\times$10$^{-4}$, whereas \object{CRL618} and \object{CRL2688} have
ratios of 6$\times$10$^{-3}$\ and 3$\times$10$^{-2}$\ respectively
\citep{Bujarrabal_etal1994}. \object{IRAS17150-3224} has an HCN/CO
ratio of 2$\times$10$^{-3}$, which is again smaller than the
2$\times$10$^{-2}$\ of \object{OH231.8+4.2}
\citep{Bujarrabal_etal1994}.

From what can be implied by Table~\ref{litabunds},
\object{IRAS17150-3224} is similar to \object{OH231.8+4.2} in
$^{13}$CO, and HCN (despite the spread in observed values). The
relatively high fractional abundance of HCN (half that of the
C-rich object \object{IRAS16594-4656}) would imply that
\object{IRAS17150-3224} is at a stage before heavy ionisation of the
molecular matter occurs (and the CN fractional abundance rises),
i.e., \object{IRAS17150-3224} is at a less evolved stage than
\object{IRAS16594-4656}.

\subsection{Chemical evolution from the AGB}
\label{chemev}

One has to ask why there seem to be two differing evolutionary paths
for PPNe: why are some PPNe molecule-rich, like \object{CRL618}, and
why are some molecule-poor, like the two objects discussed here?
Visually, \object{IRAS16594-4656} and \object{IRAS17150-3224} are not
very different from \object{CRL618}. All three have a similar bipolar
structure, with a pinched waist indicating an equatorial density
enhancement and possibly a collimating mechanism. In the case of
\object{CRL618}, this mechanism could very well be a circumstellar
torus, the presence of which has been implied from observations
\citep[see][and references therein]{HerpinCern2000}. Hence it does not
seem too exotic to consider similar structures for
\object{IRAS16594-4656} and \object{IRAS17150-3224}. Many of the PPNe
identified do have such structures, which have been directly observed
(e.g., \object{IRAS17106-3046} -- \citet{Kwok_etal2000},
\object{IRAS17245-3951} -- \citet{Hrivnak_etal1999},
\object{IRAS04296+3429} -- \citet{Sahai1999}, \object{Hen 401} --
\citet{Sahai_etal1999}, \object{IRAS17441-2411} --
\cite{Su_etal1998,Kwok_etal1996}) and there is strong evidence from
polarimetry \citep{Su_etal2000} that a circumstellar disk or torus
does exist around \object{IRAS16594-4656}.

\object{CRL618} owes its molecular richness, in a large part, to the
shielding effects of its torus. This can clearly be seen in the
chemical model of \citet{Woods_etal2003a} - the high levels of
radiation are only abated by high initial densities and a slow torus
expansion velocity, and when the density, or essentially the
extinction, becomes too low, complex molecules are destroyed. One
obvious implication that this has for PPNe such as
\object{IRAS16594-4656} and \object{IRAS17150-3224} is that their
collimating torus must not be dense enough to synthesise and protect
complex molecules. This brings in an interesting interplay between the
column density of material required to produce the degree of
collimation seen and the column density of material required to shield
complex molecules from the often very intense UV fields, cosmic rays
and possibly also X-rays. One of the prevalent results from
\citet{Woods_etal2003a} was that incident UV radiation would start to
destroy molecules when the optical depth reached approximately 10
magnitudes of extinction. This equates to a column density of
$\sim$1.6\,10$^{22}$\,cm$^{-2}$. For the particular model of
\object{CRL618}, the density at the innermost part of the torus is
$\sim$10$^9$\,cm$^{-3}$.  Numerical models
\citep[e.g.,][]{Ignace_etal1996} show that only an equatorial density
enhancement of 2 or more over the polar direction is enough to produce
asphericity (bipolarity) in a wind-compressed disk
situation. Typically an average AGB star
(e.g., $\dot{M}=10^{-5}$\,M$_\odot$\,yr$^{-1}$\ and v$_{\rm
exp}=15$\,km\,s$^{-1}$) would have a density of around
10$^7$\,cm$^{-3}$\ at a similar radial distance (10$^{15}$\,cm).

To support the hypothesis that these two PPNe are low in fractional
abundances due to low envelope density, a chemical model was
constructed to follow the evolution of a star along the AGB phase and
into the PPNe phase. Details of the model can be found in
Sect.~\ref{chemmodel}.  The model shows that within $\sim$500\,yr of
the cessation of mass-loss UV photons from the star start to ionise
the molecular material at a very high rate, and molecular matter is
destroyed rapidly. Evidence of the increasing UV field can be seen
particularly in the profiles of CN (Fig.~\ref{modelres}), which show
an increase in fractional abundance at the 500\,yr mark, where the UV
field starts to become significant. The profiles shown in
Fig.~\ref{modelres} are also subtly different from those of
steady-state AGB models (compare with the leftmost set of axes,
labelled ``0\,yr''), indicating that the mildly increasing UV field
does have a marked effect, even in the early stages of post-AGB
evolution. X-ray photons have little effect in this environment, with
the chosen assumptions. CO, which self-shields in the model, would
survive at least 12\,000\,yr after the end of mass-loss at detectable
levels. This differs from the case of a molecular torus
\citep{Woods_etal2003a} in that in the CSE case there is a reasonably
slow decline of parent species, like HCN, whereas in the torus model,
parent molecules are destroyed within a hundred years once the
extinction provided by the circumstellar matter drops to around 10
magnitudes. It is also quite marked how rapidly species form in the
torus before the \textquotedblleft radiation
catastrophe\textquotedblright, in comparison with the slow build-up in
the CSE model. As a consequence of this, there are not the high
fractional abundances of photodissociation products (e.g., CN, C$_2$H)
in the CSE case that are found in the torus model, and in
molecule-rich PPNe, like \object{CRL618}.

Using the fractional abundances calculated from observations and the
chemical model an approximate post-AGB age for \object{IRAS16594-4656}
can be estimated.  \object{IRAS16594-4656} seems to be older than
\object{IRAS17150-3224}, with quite a high CN fractional abundance and
also a higher CN/HCN ratio. Through comparison with
Fig.~\ref{modelres} in particular, the figure of 370\,yr, named by
\citet{VDV_etal1989}, seems a good estimate, since a CN/HCN ratio of
3--5 (or even 10) is only possible at times less than 500\,yr.
\object{IRAS17150-3224} is slightly more difficult to qualify, since
there is no detection of CN. However, this does suggest that the
CN/HCN ratio is low. The HCN/CO is high (higher than
\object{IRAS16594-4656}) and hence a post-AGB age of 150--210\,yr for
\object{IRAS17150-3224} \citep{Hu_etal1993,Meixner_etal2002} seems
likely, with the estimate of \citet{VDV_etal1989}, 800\,yr, appearing
to be much too long to be an accurate post-AGB age for this object.

It is not clear whether \object{IRAS16594-4656} and
\object{IRAS17150-3224} are older or younger than \object{CRL618} and
\object{CRL2688}. The lack of signs of ionisation suggest that they
are younger; the reasonably high CN/HCN ratios and low HCN/CO ratios
suggest that they are older. Certainly, \object{IRAS16594-4656} fits
into the \textquotedblleft\ evolutionary sequence\textquotedblright\
(of post-AGB objects) of \citet{Bachiller_etal1997b} between
\object{CRL618} and the young PN, \object{NGC7027}. The age estimates
of \citet{VDV_etal1989}, \citet{Hu_etal1993} and
\citet{Meixner_etal2002}, above, in comparison with the results of
\citet{Bujarrabal_etal2001} also tend to favour
\object{IRAS16594-4656} and \object{IRAS17150-3224} being older, since
\citet{Bujarrabal_etal2001} give post-AGB ages for \object{CRL618} and
\object{CRL2688} as 110 and 200\,yr, respectively.

\section{Conclusions}

Two proto-planetary nebulae, \object{IRAS16594-4656} and
\object{IRAS17150-3224}, were observed in a wide range of molecular
lines, but only detected in a few. Calculating fractional abundances
and upper limits from these lines shows that these two sources are
molecule-poor in relation to other PPNe, such as \object{CRL618},
\object{CRL2688} and \object{OH231.8+4.2}. As a reason for this
apparent difference the degree of density of the circumstellar torus
(or disk) is suggested, with molecule-rich PPNe having dense,
protective and nurturing tori, and molecule-poor PPNe having tenuous,
or no, tori. To substantiate this, a model of the chemistry in a
late--AGB/early--PPN circumstellar envelope, with no particular
density enhancements (such as a torus), is used. As expected, it shows
that very few molecules reach high fractional abundances in the
post-AGB phase, and agrees well with fractional abundances calculated
from observations. This is in stark contrast to the model of
\citet{Woods_etal2003a}, who model a dense circumstellar torus in the
PPN phase. The usefulness of HCN/CO, HNC/HCN and particularly CN/HCN
ratios in determining evolutionary age is discussed, and using these
tools, post-AGB ages for \object{IRAS16594-4656} and
\object{IRAS17150-3224} are given.

\begin{acknowledgements}
Astrophysics at UMIST is supported by a grant from PPARC. The
Swedish--ESO Submillimetre Telescope, SEST, is operated jointly by ESO
and the Swedish National Facility for Radioastronomy, Onsala Space
Observatory at Chalmers University of Technology. FLS and HO
acknowledge financial support from the Swedish research council.
\end{acknowledgements}

\bibliographystyle{aa}

\appendix
\section{Molecular line spectra}
\label{spectra}

\begin{figure*}
   \centering{   
   \includegraphics[width=17.1cm]{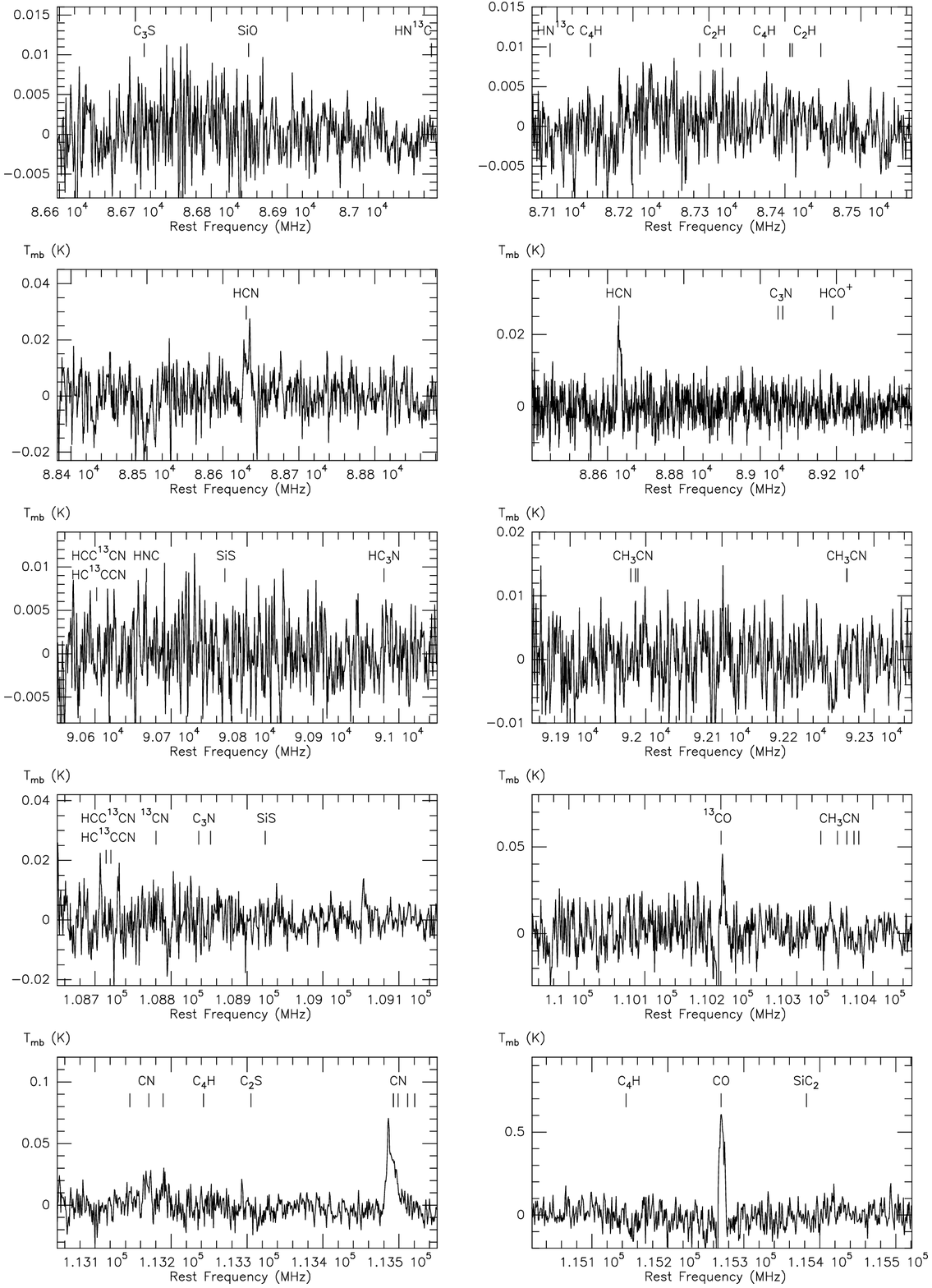}
   \caption{Molecular line observations of \object{IRAS16594-4656}.}
   \label{fig16594-1}}
   \end{figure*}

\begin{figure*}
   \centering{   
   \includegraphics[width=16.9cm]{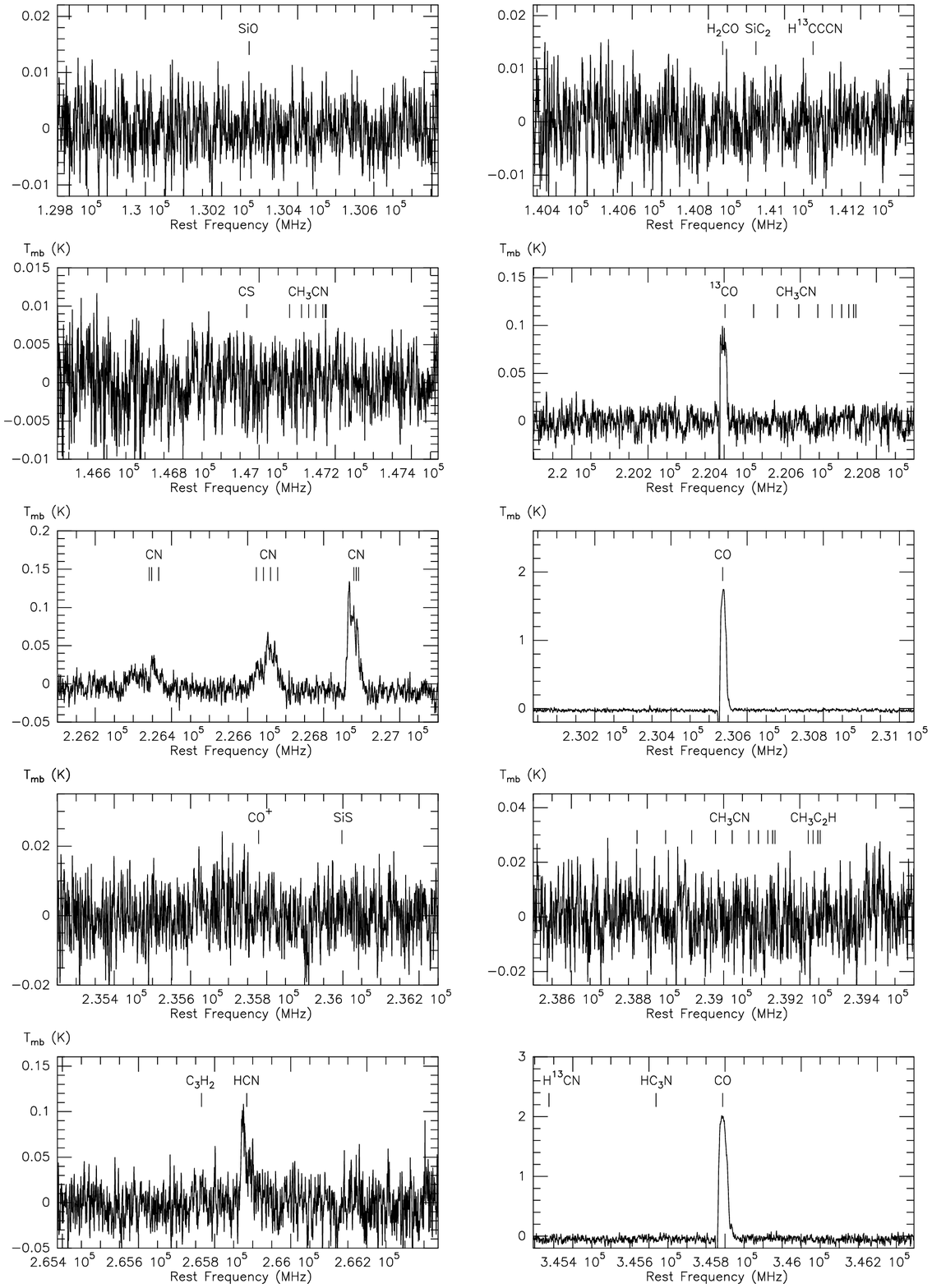}
   \caption{Molecular line observations of \object{IRAS16594-4656}.}
   \label{fig16594-2}}
   \end{figure*}

\begin{figure*}
   \centering{   
   \includegraphics[width=18cm]{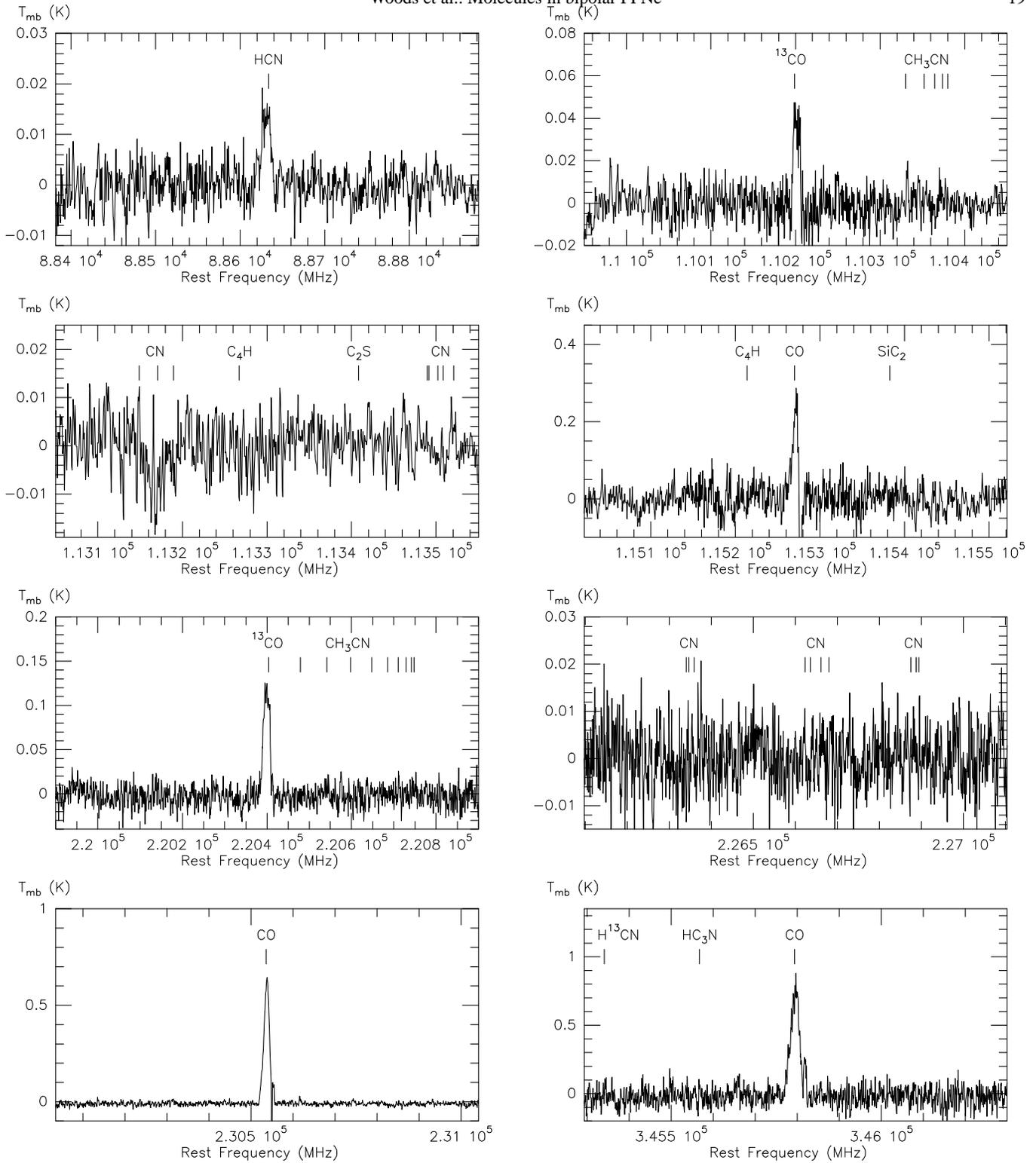}
   \caption{Molecular line observations of \object{IRAS17150-3224}.}
   \label{fig17150}}
   \end{figure*}

Spectra of \object{IRAS16594-4656} and \object{IRAS17150-3224} taken
with the SEST are presented in Figs.~\ref{fig16594-1} -- \ref{fig17150}.

\end{document}